\documentclass[12pt]{iopart}

\usepackage{cite}
\usepackage{iopams}
\usepackage{graphics} 
\usepackage{color}
\usepackage{epsfig}
\usepackage{subfigure}

\begin{document}

\title[Spin-$\frac{1}{2}$ anisotropic planar pyrochlore]
{Ground-state phase structure of the spin-$\frac{1}{2}$ anisotropic planar pyrochlore}

\author{P H Y Li and R F Bishop}
\address{School of Physics and Astronomy, Schuster Building, The University of Manchester, Manchester, M13 9PL, UK}

\begin{abstract}
  We study the zero-temperature ground-state (GS) properties of the
  spin-$\frac{1}{2}$ anisotropic planar pyrochlore, using the
  coupled cluster method (CCM) implemented to high orders of
  approximation.  The system comprises a $J_{1}$--$J_{2}$ model on the
  checkerboard lattice, with isotropic Heisenberg interactions of
  strength $J_{1}$ between
  all nearest-neighbour pairs of spins on the square lattice, and of strength $J_{2}$ between half of the next-nearest-neighbour
  pairs (in the checkerboard pattern).  We calculate
  results for the GS energy and average local GS on-site
  magnetization, using
  various antiferromagnetic classical ground states as CCM model
  states.  We also give results for the susceptibility of one of
  these states against the formation of crossed-dimer valence-bond
  crystalline (CDVBC) ordering.  The complete GS phase diagram is
  presented for arbitrary values of the frustration parameter $\kappa \equiv J_{2}/J_{1}$, and when each of
  the exchange couplings can take either sign.
\end{abstract}


\section{Introduction}
\label{intro}
The study of frustrated quantum spin systems, both theoretically and
experimentally, has become a field of intense activity in recent
years, especially in the context of quantum phase
transitions~\cite{Sachdev:1995_QM,Scholl:2004_2Dmagnetism,Misguich:2005_2D_magnetism-coll}.
In particular, the experimental investigation of a wide variety of
quasi-two-dimensional materials with fascinating properties has
progressed hand-in-hand with the theoretical study of spin-lattice
models believed to capture their most important behaviour.  The
interplay between frustration, both geometrical and dynamical, and
quantum fluctuations can produce interesting zero-temperature ($T=0$)
ground-state (GS) quantum phase transitions.  These can involve phases
ranging from ones with quasiclassical magnetic long-range order (LRO), such
as N\'{e}el antiferromagnetic (AFM) order, to others such as
valence-bond solids and spin liquids, which have no classical
counterparts.  The stable GS phases are strongly influenced by
parameters such as the spin quantum number $s$ of the magnetically
active ions situated on the sites of the lattice, the dimensionality,
the coordination number ($z$), and geometry of the lattice, the number
of magnetic bonds and whether they are ferromagnetic (FM) or AFM in
nature.

One of the key motivators that has spurred continued interest in
highly frustrated quantum spin-lattice systems, has been the
possibility of finding situations where a novel phase with no
classical counterpart forms the stable GS phase.  In particular, the
search for quantum spin-liquid phases has received intense interest,
ever since they were first proposed by Anderson
\cite{Anderson:1973_QSL} over 40 years ago.  Even though we now know that
the example proposed by Anderson, namely, the GS phase of a spin-1/2
Heisenberg antiferromagnet (HAFM) on a triangular lattice (i.e., with
spins interacting via nearest-neighbour (NN) isotropic Heisenberg AFM
exchange interactions only), is not a spin liquid, the search for such
states continues in other systems.  It is widely believed that prime
candidates in this context are those which at the classical ($s
\rightarrow \infty$) level display a GS phase in some region of their
phase diagram that has macroscopic degeneracy.  In particular, much
attention has thereby focused on frustrated quantum spin systems that
are built from tetrahedra coupled into two-dimensional (2D) or
three-dimensional (3D) lattice networks
\cite{Singh:1998_chkboard,Canal:1998_chkboard,Palmer:2001_chkboard,Brenig:2002_chkboard,Canals:2002_chkboard,Starykh:2002_chkboard,Sindzingre:2002_chkboard,Fouet:2003_chkboard,Fukazawa:2003_chkboard,Berg:2003_chkboard,Tchernyshyov:2003_chkboard,Moessner:2004_chkboard,Hermele:2004_chkboard,Brenig:2004_chkboard,Bernier:2004_chkboard,Starykh:2005_chkboard,Schmidt:2006_chkboard,Arlego:2007_chkboard,Arlego:2009_chkboard,Moukouri:2008_chkboard,Chan:2011_chkboard,Bishop:2012_checkerboard}.

\begin{figure*}[!tb]
\vskip0.2cm
\begin{center}
\mbox{
\subfigure[]{\scalebox{0.35}{\includegraphics{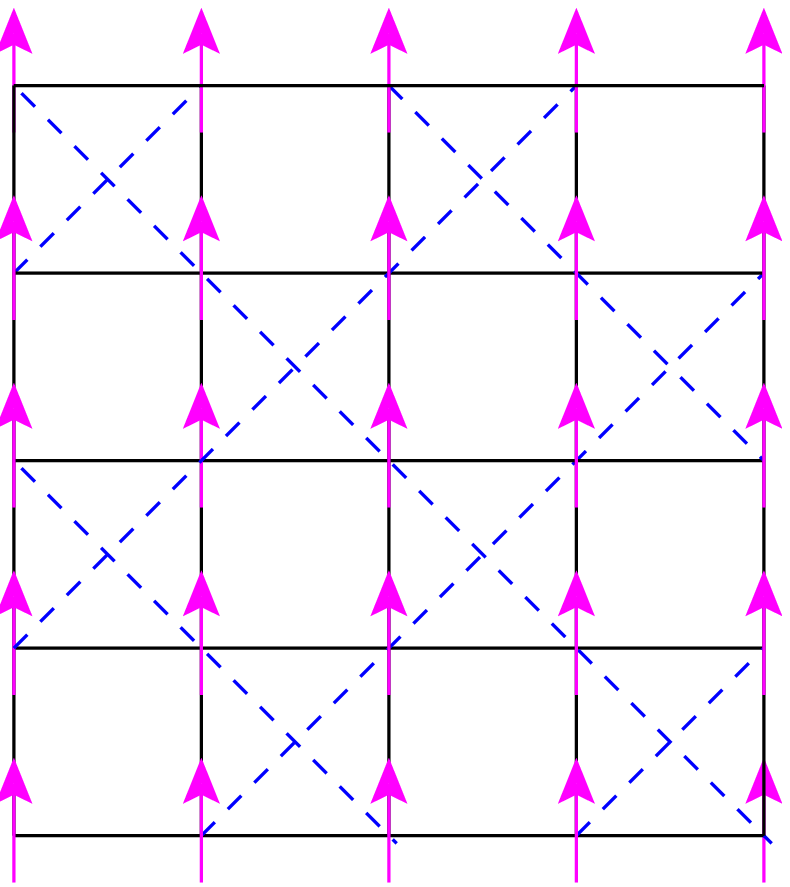}}} \quad 
\subfigure[]{\scalebox{0.35}{\includegraphics{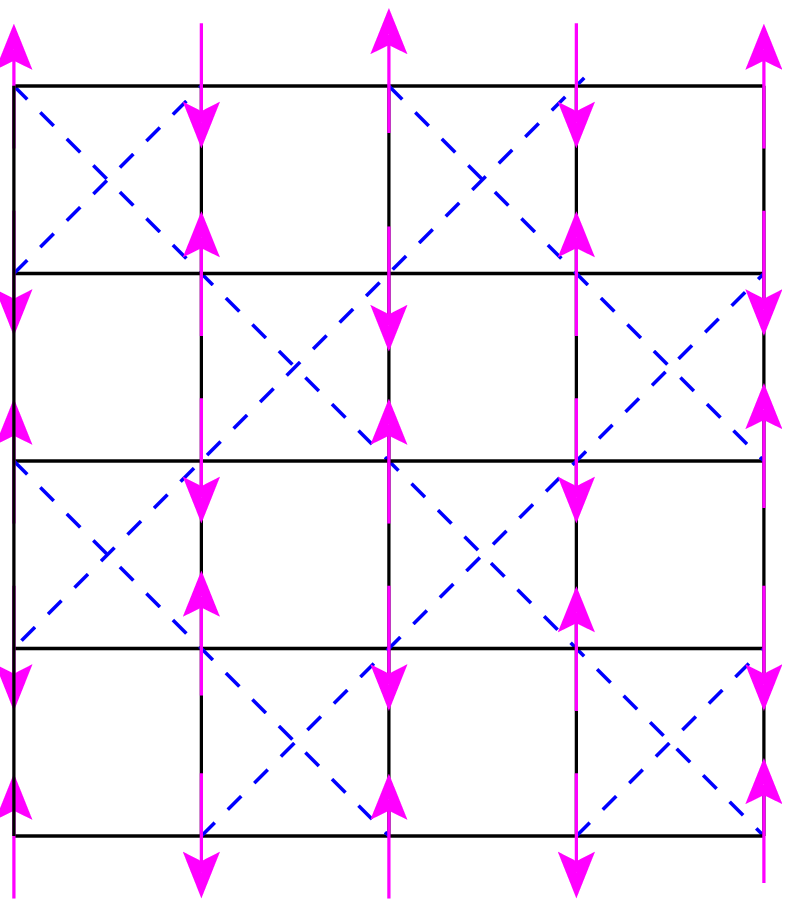}}} \quad
\subfigure[]{\scalebox{0.35}{\includegraphics{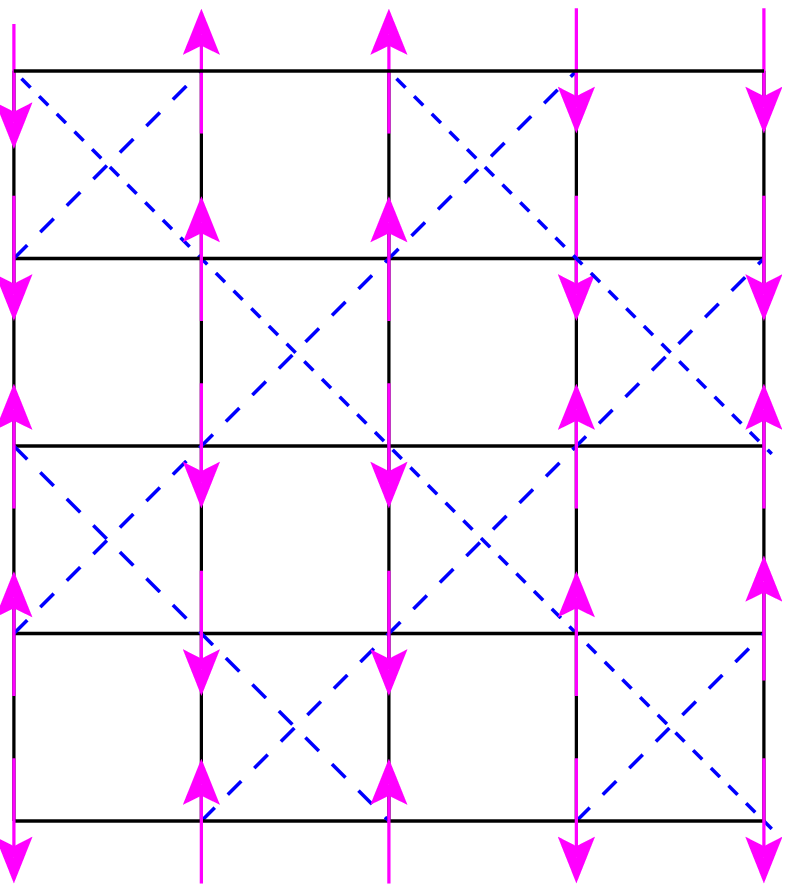}}} \quad
\subfigure[]{\scalebox{0.35}{\includegraphics{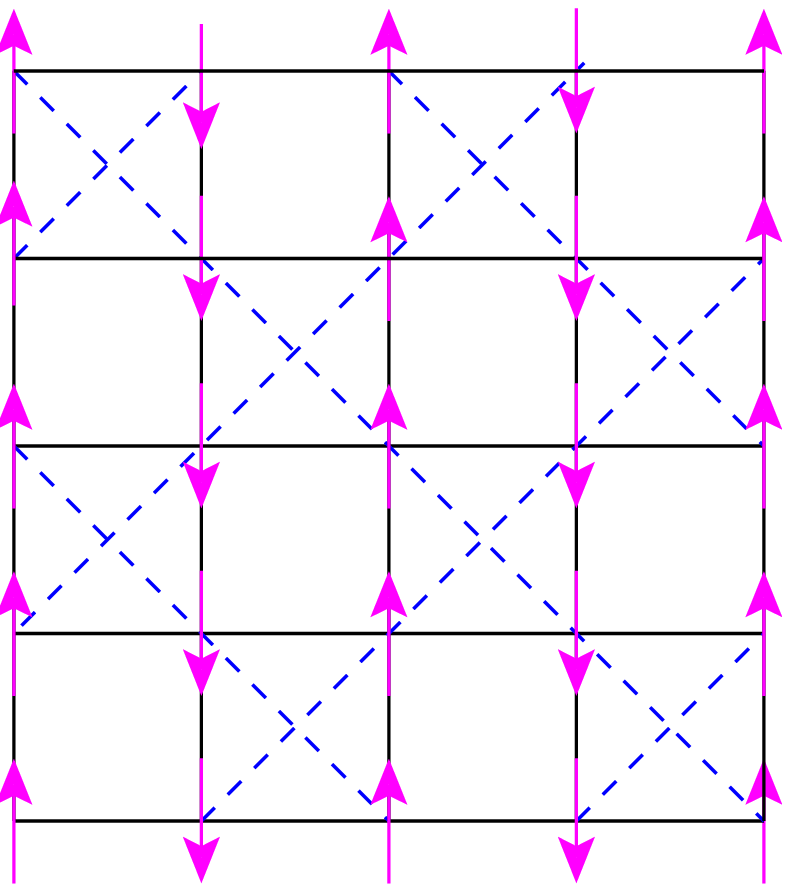}}} 
}
\vskip0.9cm
\mbox
{
\subfigure[]{\scalebox{0.5}{\includegraphics[angle=90]{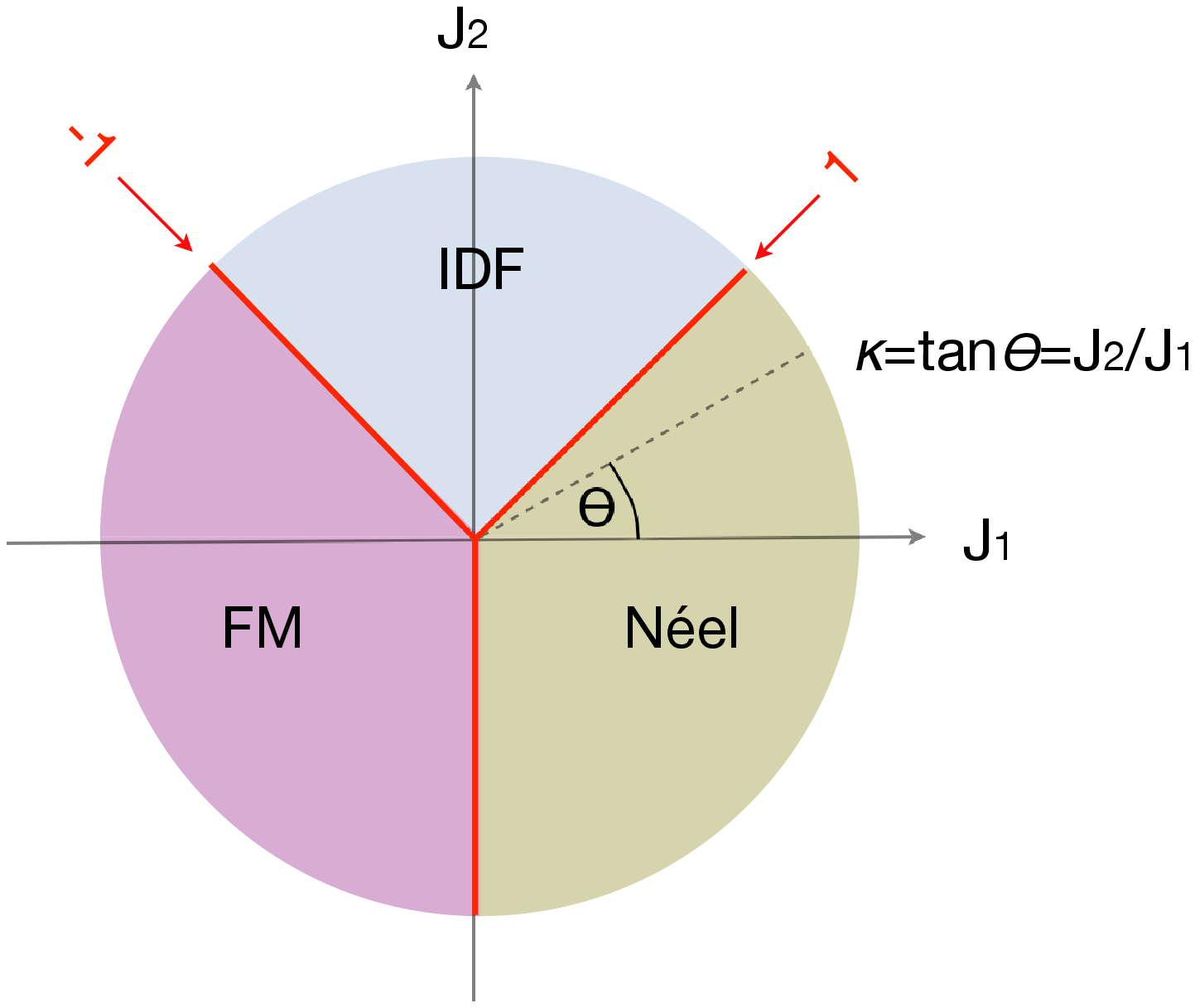}}}
}
\caption{The $J_{1}$--$J_{2}$ Heisenberg model on the checkerboard lattice, showing (a) the ferromagnetic (FM) state, (b) the N\'{e}el state, (c) one of the two
  N\'{e}el$^{\ast}$ states, and (d) one (viz., the columnar) of the two striped states.  The solid (black) lines are NN $J_{1}$
  bonds and dashed (green) lines are NNN $J_{2}$ bonds.  The spins on
  each lattice site are portrayed by the (brown) arrows.  (e) The classical $T=0$ phase diagram.}
\label{model_bonds}
\end{center}
\end{figure*}

The pyrochlore lattice comprises a 3D arrangement of corner-sharing
tetrahedra, and it is well known that essentially all compounds that
crystallize into the pyrochlore structure display unusual magnetic
properties.  In order to simplify the study of the 3D pyrochlore, but
without losing any of its magnetic frustration, it is also common to
project the 3D vertex-sharing lattice of tetrahedra onto a 2D plane.
Each tetrahedron has four spins at its vertices, with each of its six
edges or links symbolizing a Heisenberg interaction.  In the 2D
projection (viz., the planar pyrochlore) each tetrahedron is mapped to
a square with spins at its vertices, with the sides denoting the NN
bonds with Heisenberg exchange bonds of coupling strength $J_{1}$,
plus additional exchange bonds of coupling strength $J_{2}$ across its
diagonals, i.e., now joining next-nearest-neighbour (NNN) bonds in
the square-lattice geometry.  Each such square is then coupled to
others of the same kind by NN $J_{1}$ bonds, resulting in the
checkerboard pattern shown in figure \ref{model_bonds}.  

The isotropic
planar pyrochlore is simply the case $J_{2}=J_{1}$, but it is of
considerable interest to study also the anisotropic case when $J_{2}
\neq J_{1}$.  While the 2D projection of the 3D pyrochlore pattern preserves the
vertex-sharing structure, nevertheless one loses the symmetry between
the six equivalent bonds on each tetrahedron in the 3D structure when
the 2D projection is made.  Thus, in the planar pyrochlore the two
diagonal bonds of each crossed square are inequivalent to the four
bonds on the sides of the square.  This subsequent symmetry reduction
is thus itself completely consistent with considering the case
$J_{2} \neq J_{1}$ of the anisotropic Heisenberg model on the 2D
checkerboard lattice or, equivalently, the anisotropic planar
pyrochlore.  Alternate names are the crossed chain model and the
$J_{1}$--$J_{2}$ model on the checkerboard lattice.

Most interest to date on the anisotropic planar pyrochlore has
focussed on the case when both bonds are AFM in nature (i.e.,
$J_{1}>0,\,J_{2}>0$).  On the other hand, there has also been
increasing interest at both the experimental and theoretical levels in
quasi-2D magnetic material in which the NN coupling is ferromagnetic
($J_{1}<0$), while NNN coupling is antiferromagnetic ($J_{2} > 0$).
Examples of such materials include Pb$_{2}$VO(PO$_{4}$)$_{2}$
\cite{Kaul:2004_FM,Skoulatos:2007_FM,Carretta:2009_FM,Skoulatos:2009_FM,Nath:2009_FM},
(CuCl)LaNb$_{2}$O$_{7}$ \cite{Kageyama:2005_FM},
SrZnVO(PO$_{4}$)$_{2}$
\cite{Skoulatos:2009_FM,Tsirlin:2009_PRB79_FM,Tsirlin:2009_PRB80_FM,Nath:2008_FM},
BaCdVO(PO$_{4}$)$_{2}$
\cite{Carretta:2009_FM,Tsirlin:2009_PRB80_FM,Nath:2008_FM},
PbZnVO(PO$_{4}$)$_{2}$ \cite{Tsirlin:2010_FM} and, (CuBr)LaNb$_2$O$_7$
\cite{oba2006_FM}.  Experimental studies of these and other materials
have, in turn, stimulated theoretical interest in the GS and
thermodynamic properties of the $J_{1}$--$J_{2}$ model on the square
lattice when the NN exchange bond is FM in nature ($J_{1}<0$) and the
NNN exchange bond is AFM, and hence frustrating, in nature ($J_{2}>0$)
{\cite{Shannon:2004_FM,Shannon:2006_FM,Sindzingre:2007_FM,Schmidt:2007_JPCM_FM,Schmidt:2007_JMMM_FM,Viana_Sousa:2007_FM,
    Sindzingre_Shannon:2009_FM,Sindzingre:2009_FM,Hartel:2010_FM,Richter2010:J1J2mod_FM,Nunes_Sousa:2011_FM,Shindou_Momoi:2011_FM,Momoi:2006_FM}.
  It is of particular interest to note in this context that there has
  been a considerable degree of controversy between various
  theoretical studies on whether or not a magnetically disordered,
  spin-nematic phase emerges in the spin-$\frac{1}{2}$ model between the phases
  exhibiting quasiclassical collinear striped AFM order and FM order.
  In a similar vein we have also recently studied ourselves the
  spin-$\frac{1}{2}$ $J_{1}$--$J_{2}$--$J_{3}$ model on the honeycomb lattice
  with FM NN bonds (of strength $J_{1}<0$) and AFM NNN and
  next-next-nearest-neighbour bonds of equal strength
  ($J_{3}=J_{2}>0$) \cite{PHYLi:2012_honeycomb_J1neg}.

  Given the high level of interest in the frustrated spin-$\frac{1}{2}$
  $J_{1}$--$J_{2}$ Heisenberg ferromagnet on the square lattice (i.e.,
  with $J_{1}<0,\,J_{2}>0$) noted above, it seems timely to consider
  now the analogous model on the checkerboard lattice.  In a recent
  paper \cite{Bishop:2012_checkerboard} we studied the frustrated
  spin-$\frac{1}{2}$ HAFM on the checkerboard lattice (i.e., with $J_{1}>0,
  J_{2}>0$), using the coupled cluster method (CCM).  Our aim now is
  to extend that analysis to investigate the entire $T=0$ GS phase
  diagram, when both exchange couplings $J_{1}$ and $J_{2}$ can take
  either sign.  After describing and discussing the model itself in
  section \ref{model}, we discuss the CCM methodology in section
  \ref{ccm}.  Our results are then presented in section \ref{results},
  and we end with a summary and conclusions in section \ref{summary}.

\section{The model}
\label{model}
The Hamiltonian for the anisotropic planar pyrochlore is that of a $J_{1}$--$J_{2}$ model on a checkerboard lattice.  It is written as
\begin{equation}
H = J_{1}\sum_{\langle i,j \rangle} \mathbf{s}_{i}\cdot\mathbf{s}_{j} + J_{2}\sum_{\langle\langle i,k \rangle\rangle} 
\mathbf{s}_{i}\cdot\mathbf{s}_{k}\,, \label{H}
\end{equation}
where the index $i$ runs over all $N$ sites of a square lattice; $j$
runs over all four NN sites to site $i$ that are
connected to it by $J_{1}$ bonds; and $k$ runs over only two of the
NNN sites to site $i$ that are connected to
it by $J_{2}$ bonds in the checkerboard pattern shown in figure
\ref{model_bonds}.  We are interested in the thermodynamic limit ($N
\rightarrow \infty$) of an infinite lattice.  The sums on $\langle i,j
\rangle$ and $\langle \langle i,k \rangle \rangle$ count each pairwise
bond once and once only.  Thus, the alternate square plaquettes of the
checkerboard model have either two NNN (diagonal) bonds (i.e., a
cross) or none, as shown in figure \ref{model_bonds}.  Each site $i$
of the lattice carries a particle with a spin quantum number $s$,
defined by a spin operator ${\bf s}_{i} =
(s^{x}_{i},s^{y}_{i},s^{z}_{i})$.
The checkerboard model is also known as the crossed chain model, which
originates from its crossed (in our case, diagonal in figure
\ref{model_bonds}) sets of chains, with $J_{2}$ the intrachain
exchange coupling constant, and $J_{1}$ the (in our case, vertical and
horizontal) interchain exchange bonds that connect NN pairs of spins
from chains running across one another.  We will be interested here in
the case $s=1/2$, but it is worthwhile first to discuss briefly the
classical limit ($s \rightarrow \infty$) of the model, in order to
compare with the effects caused by quantum fluctuations.

For $J_{2}=0$ the classical GS phase is the FM state
shown in figure \ref{model_bonds}(a) for $J_{1}<0$ and the AFM
N\'{e}el state shown in figure \ref{model_bonds}(b) for $J_{1}>0$.
Both states are unique for specified values of the ordering vectors.
The N\'{e}el state has every row and column joined by $J_{1}$ bonds
with NN spins oriented antiparallel to one another, and hence with
each of the diagonal crossed $J_{2}$-chains ordered ferromagnetically
with all of the spins on a given diagonal oriented parallel to one
another, but with spins on diagonals in a given direction oriented
antiparallel to those on the diagonals in the perpendicular direction.
Both the N\'{e}el and the FM states will clearly remain the stable GS
phases in their respective domains for sufficiently small values of
$|J_{2}|$.

The GS energy of the classical N\'{e}el state is thus $E^{{\rm
    cl}}_{{\rm N\acute{e}el}}/N = s^{2}(-2J_{1}+J_{2})$.  Clearly, for
sufficiently large values of $J_{2}>0$ the spins on the diagonals (or
crossed $J_{2}$-chains) will prefer to align antiferromagnetically.
It is trivial to see that there is an infinitely degenerate family
(IDF) of collinear AFM phases in which every diagonal $J_{2}$-chain
displays N\'{e}el AFM ordering, but where every diagonal (each of
which is connected by $J_{1}$ bonds to two other crossed diagonal
$J_{2}$-chains) can be arbitrarily displaced along its own direction
or, equivalently, where all of its spins have their directions
reversed.  All of these states have the same classical energy per spin
given by $E^{{\rm cl}}_{{\rm IDF}}/N = -s^{2}J_{2}$, completely
independent of the exchange coupling $J_{1}$.

When $J_{1}>0$ there is thus only one classical phase transition in
the model, at $\kappa^{a}_{{\rm cl}}=1$, between the N\'{e}el AFM
phase and the IDF of AFM phases, where $\kappa \equiv J_{2}/J_{1}$ is the frustration (or planar pyrochlore anisotropy) parameter.  One particular member of this IDF is
the so-called N\'{e}el$^{\ast}$ state
\cite{Bishop:2012_checkerboard}, (one of which is) shown in
figure \ref{model_bonds}(c).  It has doubled AFM order, $\cdots
\uparrow \uparrow \downarrow \downarrow \uparrow \uparrow \downarrow
\downarrow \cdots$, along each row and column of sites joined by
$J_{1}$ bonds, such that the single-site $\uparrow$ or $\downarrow$
spin in the N\'{e}el state is replaced by the two site $\uparrow
\uparrow$ or $\downarrow \downarrow$ unit in the N\'{e}el$^{\ast}$
state.  The N\'{e}el$^{\ast}$ state is actually doubly degenerate (for
a given direction of the N\'{e}el vector), since a translation along a
lattice (square-plaquette) diagonal [i.e., a translation of one
lattice vector in the horizontal direction plus one lattice vector in
the vertical direction of figure \ref{model_bonds}(c)] transforms one
of the pair into the other.  Two other simple members of the IDF of
classical GS phases for $\kappa > 1$ (and $J_{1}>0$) are the so-called
columnar and row striped states \cite{Bishop:2012_checkerboard}, the former of which is shown in figure \ref{model_bonds}(d).
These have, respectively, columns or rows of spins joined by $J_{1}$
bonds ordered in a FM fashion, with spins on alternating columns or
rows oriented antiparallel to one another, and hence with N\'{e}el AFM
ordering along rows or columns, respectively, for the two cases.  Like
the N\'{e}el$^{\ast}$ state, the striped states are doubly degenerate
(for a given N\'{e}el vector), since the row and columnar states
transform into one another under interchange of rows and columns.

The GS energy of the classical FM state is $E^{{\rm cl}}_{{\rm FM}}/N
= s^{2}(2J_{1}+J_{2})$, and hence there is a second classical phase
transition at $\kappa^{b}_{{\rm cl}}=-1$ between the FM state and the
IDF of AFM states, when $J_{1}<0$.  Finally, there is a third
classical phase transition between the N\'{e}el AFM state and the FM
state at $J_{1}=0$, when $J_{2}<0$ (or, equivalently at
$\kappa^{c}_{{\rm cl}}=\pm \infty$, with $J_{2}<0$).  In summary we
have three classical GS phases, namely (a) the N\'{e}el AFM state in
the regime $J_{1}>0$ and $-\infty < J_{2} < J_{1}$, (b) the IDF of AFM
states for $J_{2} > |J_{1}|$, and (c) the FM state for $J_{1}<0$ and
$-\infty < J_{2} < -J_{1}$, all as shown in figure \ref{model_bonds}(e).

Clearly the model has several interesting special limiting cases.  For
example, in the case of AFM NN couplings ($J_{1}>0$), the model
reduces to the isotropic Heisenberg antiferromagnet (HAFM) on the
square lattice as $\kappa \rightarrow 0$, and to decoupled
one-dimensional (1D) crossed isotropic HAFM chains as $\kappa
\rightarrow \infty$.  In between, at $\kappa=1$, the model is simply
the isotropic HAFM on the checkerboard lattice, which is itself a 2D
analogue of the 3D isotropic pyrochlore HAFM.  Compared to the
classical version ($s \rightarrow \infty$) of the model, for which the
full GS phase diagram in the $J_{1}$--$J_{2}$ plane is known, the
$s=1/2$ case is basically only well established at precisely the above
three points when $J_{1}>0$.

Thus, for the spin-$\frac{1}{2}$ HAFM on the square lattice
($\kappa=0$) there is a general consensus that the classical N\'{e}el
AFM LRO is weakened but not destroyed by quantum fluctuations, such
that the sublattice magnetization is reduced to about 61.5\% of its
classical value of 0.5 for the $s=1/2$ case
\cite{Zheng:1991_SqLatt,Beard:1996_SqLatt,Sandvik:1997_SqLatt,Sandvik:2010_SqLatt,DJJFarnell:2014_archimedeanLatt}.
The excited states are also well established to be gapless
integer-spin magnons for the $s=1/2$ model.  One then expects, by
continuity, that this (partial) N\'{e}el AFM order will persist as the
frustrating NNN bonds (with strength $J_{2}>0$) are turned on.  As
$J_{2}$ increases (with $J_{1}>0$ held fixed) we expect that this
order will ultimately vanish at some critical value $\kappa =
\kappa_{c_{1}}$, where the N\'{e}el staggered magnetization becomes
zero.  In the opposite limit, $\kappa \rightarrow \infty$ (i.e.,
$J_{2} \rightarrow \infty$ with $J_{1}>0$), where we have decoupled 1D
isotropic HAFM chains, the model is exactly soluble for the
spin-$\frac{1}{2}$ case.  The GS phase is a Luttinger spin liquid, in
which the quantum fluctuations have completely destroyed the classical
N\'{e}el LRO, such that the N\'{e}el staggered magnetization vanishes.
The excited states are gapless, deconfined, spin-$\frac{1}{2}$
spinons.

In between these two limits, at $\kappa=1$, most studies now concur (and see, e.g., \cite{Fouet:2003_chkboard,Tchernyshyov:2003_chkboard,Brenig:2004_chkboard,Starykh:2005_chkboard,Moukouri:2008_chkboard,Chan:2011_chkboard,Bishop:2012_checkerboard})
that the GS phase of the $s=1/2$ isotropic HAFM on the checkerboard
lattice is a plaquette valence-bond crystal (PVBC) with quadrumer LRO
on isolated spin-singlet square plaquettes, on top of which there is
an excitation spectrum of gapped, confined, integer-spin spinons.
Again, by continuity, one expects that this PVBC order will persist
over a (finite) range of the anisotropy parameter $\kappa$ on
either side of the isotropic value $\kappa=1$.

On the low-$\kappa$ side it is still not completely settled whether
there is a direct (i.e., first-order in the Landau-Ginzburg scenario)
phase transition between the states with N\'{e}el AFM order and PVBC
order at $\kappa=\kappa_{c_{1}}$, or, if not, there exists an
intermediate coexistence phase with both types of ordering and, hence,
with two different order parameters.  On the high-$\kappa$ side we
expect that PVBC order will persist out to some critical value
$\kappa_{c_{2}}>1$.  The situation for $\kappa>\kappa_{c_{2}}$ (with
$J_{1}>0$) has been, up till now, the most unsettled part of the phase
diagram.  Various scenarios have been put forward, as have been
summarized in our own earlier paper \cite{Bishop:2012_checkerboard}.
In particular, it is an obvious question to ask if any of the IDF of
classical ($s \rightarrow \infty$) AFM states that exist for $\kappa
> \kappa^{a}_{{\rm cl}}=1$, survives the quantum fluctuations present
in the $s=1/2$ model.  Furthermore, if the answer is affirmative, one
may then enquire as to whether the classical degeneracy might be
lifted by the well-known ``order-by-disorder'' mechanism
\cite{Villain:1977_ordByDisord, Villain:1980_ordByDisord}.

In a recent previous paper \cite{Bishop:2012_checkerboard} we studied
the $s=1/2$ model on the anisotropic checkerboard lattice, using the
CCM, in the case when both NN and NNN exchange couplings are AFM in
nature (i.e., with $J_{1}>0$, $J_{2}>0$).  We used various AFM
classical ground states as CCM reference states, including the
N\'{e}el state of figure \ref{model_bonds}(b), the N\'{e}el$^{\ast}$
state of figure \ref{model_bonds}(c), and the (columnar)
striped state of figure \ref{model_bonds}(d).  We thereby presented results for the
GS energy and average local on-site magnetization (i.e., the magnetic
order parameter) of these states, including their susceptibilities
against the formation of both PVBC order and crossed-dimer valence
bond crystal (CDVBC) order.  The CDVBC state is one with dimer LRO on
isolated spin-singlet dimers arranged in a pattern of crossed dimers
on alternating square plaquettes (i.e., on every second row and every
second column).  Our main findings are summarized as follows.

Firstly, we showed that the AFM quasiclassical state with N\'{e}el ordering
is indeed the GS phase for $\kappa<\kappa_{c_{1}} \approx 0.80 \pm
0.01$.  Secondly, we showed that although quantum fluctuations do
indeed lift the classical degeneracy of the IDF of AFM states that
form the GS phase for the classical version ($s \rightarrow \infty$) of
the model for $\kappa > 1$, with the N\'{e}el$^{\ast}$ states having
a lower energy than the striped states, all of these states are
actually magnetically unstable in the sense that their magnetic order
parameters are zero (or negative) within very small numerical errors.
Thirdly, we showed that our calculations preferred instead a
PVBC-ordered phase for $\kappa_{c_{1}} < \kappa < \kappa_{c_{2}}
\approx 1.22 \pm 0.02$, and a CDVBC-ordered phase for {\it all} values of
$\kappa > \kappa_{c_{2}}$.  Lastly, our calculations indicated that
both transitions (i.e., from N\'{e}el to PVBC, and from PVBC to CDVBC)
are probably direct ones.  Nevertheless, we could not entirely rule
out regions of coexistence in both cases, although we showed that if
they do exist they must be very narrow, being confined respectively to
$0.79 \lesssim \kappa \lesssim 0.81$ and $1.20 \lesssim \kappa
\lesssim 1.22$.

The main purpose of the present paper is now to extend the above
analysis to the entire phase space, where both NN and NNN exchange
bonds can now be either AFM or FM in nature.  A particular aim will be
to examine the phase boundary of the CDVBC state when the NN bonds are
allowed to become FM (i.e, with $J_{1}<0$).

\section{The coupled cluster method}
\label{ccm}
The CCM is a universal method of {\it ab initio} quantum many-body
theory, which is open to systematic step-by-step improvements via
various well-defined approximation schemes.  It has been applied with
considerable success to a wide spectrum of both finite and extended
physical systems defined either in a spatial continuum or on a regular
discrete lattice \cite{Bishop:1987_ccm,Bartlett:1989_ccm,Arponen:1991_ccm,Bishop:1991_XXZ_PRB44,Bishop:1991_TheorChimActa_QMBT,Bishop:1998_QMBT_coll,Fa:2004_QM-coll}.
These range from atoms and molecules of interest in quantum chemistry
to the electron gas; from closed- and open-shell atomic nuclei to infinite nuclear matter;
from various strongly interacting quantum field theories to models of
interest in quantum optics, quantum electronics and solid-state
optoelectronics; as well as to many condensed-matter systems, including
highly-frustrated and strongly-correlated spin-lattice systems of the
kind considered here.

The CCM is particularly appropriate for such quantum magnets, for
which many of the alternative methods have serious drawbacks.  For
example, quantum Monte Carlo (QMC) techniques are hindered by the
well-known minus-sign problem, which is usually unavoidable for
frustrated spin-lattice problems.  Similarly, the exact
diagonalization (ED) method is usually confined to such relatively
small finite-lattice clusters that the more subtle orderings present
in the GS phase diagram may be difficult to detect and calculate
accurately.  The CCM suffers from neither of these problems, and has
been applied with great success in recent years to many different
spin-lattice models of topical interest in quantum magnetism (and see,
e.g., \cite{Bishop:2012_checkerboard,Bishop:1991_XXZ_PRB44,Fa:2004_QM-coll,RoHe:1990_1D-2D,Zeng:1998_SqLatt_TrianLatt,SEKruger:2006_spinStiff,Darradi:2008_J1J2mod,Bi:2009_SqTriangle,Richter2010:J1J2mod_FM,Gotze:2011_kagome,PHYLi:2012_honeycomb_J1neg,Bishop:2013_crossStripe,DJJFarnell:2014_archimedeanLatt,Richter:2015_ccm_J1J2sq_spinGap} and references cited therein).

Since the CCM methodology has been described in detail elsewhere (and
see, e.g., \cite{Arponen:1991_ccm,Bishop:1991_XXZ_PRB44,Bishop:1991_TheorChimActa_QMBT,Bishop:1998_QMBT_coll,Fa:2004_QM-coll,Zeng:1998_SqLatt_TrianLatt,Bi:2009_SqTriangle,Bishop:2013_crossStripe,DJJFarnell:2014_archimedeanLatt}),
we focus only on its key elements here.  We note that the method is
size-extensive, and can hence provide results in the thermodynamic
limit ($N \rightarrow \infty$) from the outset.  The first step is
always to choose a suitable, normalized $N$-body model (or reference)
state, $|\Phi\rangle$, on top of which the quantum fluctuations
present in the exact GS wave function $|\Psi\rangle$ of the phase
under study can be incorporated at the next stage, described below.
The conditions that $|\Phi\rangle$ must satisfy are also described below.
The exact GS ket- and bra-state wave functions, which satisfy the
respective Schr\"{o}dinger equations,
\begin{equation}
H|\Psi\rangle=E|\Psi\rangle\,; \quad \langle\tilde{\Psi}|H = E\langle\tilde{\Psi}|\,,  \label{schrodinger_eq}
\end{equation}
are chosen to have normalizations such that
$\langle\tilde{\Psi}|\Psi\rangle = \langle{\Phi}|\Psi\rangle =
\langle{\Phi}|\Phi\rangle = 1$.  

A key element of the CCM is then to parametrize these exact states in
terms of the chosen model state via the distinctive exponentiated
forms,
\begin{equation}
|\Psi\rangle=e^{S}|\Phi\rangle\,; \quad \langle\tilde{\Psi}|=\langle\Phi|\tilde{S}e^{-S}\,.  \label{exp_para}
\end{equation}
In turn, the two correlation operators, $S$ ($\tilde{S}$), are
themselves formally decomposed in terms of a mutually commuting set of
$N$-body creation operators $\{C^{+}_{I}\}$ (annihilation operators
$\{C^{-}_{I} \equiv (C^{+}_{I})^{\dagger}\}$), as
\begin{equation}
S=\sum_{I\neq 0}{\cal S}_{I}C^{+}_{I}\,; \quad \tilde{S}=1+\sum_{I\neq 0}\tilde{{\cal S}}_{I}C^{-}_{I}\,,  \label{correlation_oper}
\end{equation}
where we define $C^{+}_{0}\equiv 1$ to be the identity operator, and
where the set index $I$ denotes a complete set of single-particle
configurations for all $N$ particles.  The model state $|\Phi\rangle$
and the operators $\{C^{+}_{I}\}$ must be selected so that
$|\Phi\rangle$ is a fiducial (or cyclic) vector with respect to the
set of mutually commuting creation operators $\{C^{+}_{I}\}$.  The
model state thus plays the role of a generalized vacuum state, with
$\langle\Phi|C^{+}_{I} = 0 = C^{-}_{I}|\Phi\rangle\,, \forall I \neq
0$.  Furthermore the set of states $\{C^{+}_{I}|\Phi\rangle\}$ spans
the entire Hilbert space to which $|\Psi\rangle$ belongs.

For applications to spin-lattice problems it is convenient to consider
each lattice site $k$ in the chosen model state $|\Phi\rangle$ to be
equivalent to all others.  A simple way to do so is to rotate
(passively) each spin on each site $k$ separately in such a way that
every spin points downward (say along the negative $z$-direction) in
its own local frame of spin axes.  Such choices of local spin
coordinates leave the basic SU(2) spin commutation relations
unchanged.  However, all independent-spin product model states thereby
take the universal form $|\Phi\rangle =
|\downarrow\downarrow\downarrow\cdots\downarrow\rangle$, and the
$N$-body creation operators $C^{+}_{I}$ similarly take a universal
product form, $C^{+}_{I} \equiv s^{+}_{k_{1}}s^{+}_{k_{2}}\cdots
s^{+}_{k_{n}}$, in terms of the single-spin raising operators,
$s^{+}_{k} \equiv s^{x}_{k}+is^{y}_{k}$.  If each site carries a spin
with spin quantum number $s$, no site-index $k_{j}$ in the product form for
$C^{+}_{I}$ may appear more than $2s$ times (i.e., for the present
$s=\frac{1}{2}$ case, each index may appear at most once).  Clearly
the set index $I$ thus simply becomes $I \equiv \{k_{1},k_{2},\cdots ,
k_{n};\; n=1,2,\cdots , 2sN\}$.  With the choice of local spin
coordinates thus made separately for each model state, the Hamiltonian
simply needs to be suitably rewritten in terms of the selected
spin-coordinate frames.

The CCM correlation coefficients $\{{\cal S}_{I}, \tilde{{\cal
    S}}_{I}\}$, which completely determine all GS properties, are now
themselves calculated by minimization of the GS energy expectation
value functional,
\begin{equation}
\bar{H}=\bar{H}[{\cal S}_{I},{\tilde{\cal S}_{I}}]\equiv
\langle\Phi|\tilde{S}e^{-S}He^{S}|\Phi\rangle\,,  \label{GS_E_xpect_funct}
\end{equation}
with respect to each of the coefficients $\{{\cal S}_{I},{\tilde{\cal
    S}}_{I}\,; \forall I \neq 0\}$.  Variation of equation
(\ref{GS_E_xpect_funct}) with respect to $\tilde{\cal S}_{I}$ from
equation (\ref{correlation_oper}) yields the coupled set of
non-linear equations,
\begin{equation}
\langle\Phi|C^{-}_{I}e^{-S}He^{S}|\Phi\rangle = 0\,, \quad \forall I \neq 0\,,  \label{ket_eq}
\end{equation}
for the set of creation correlation coefficients $\{{\cal S}_{I}\}$.  Similarly, variation of equation
(\ref{GS_E_xpect_funct}) with respect to ${\cal S}_{I}$ from
equation (\ref{correlation_oper}) yields the
corresponding set of linear equations,
\begin{equation}
\langle\Phi|\tilde{S}e^{-S}[H,C^{+}_{I}]e^{S}|\Phi\rangle\,, \quad \forall I \neq 0\,,  \label{bra_eq}
\end{equation}
for the set of annihilation correlation coefficients $\{{\tilde{{\cal S}}}_{I}\}$, once equation (\ref{ket_eq})
has first been solved for the coefficients $\{{\cal S}_{I}\}$.  Since $[S, C^{+}_{I}]=0$ by construction, we note that equation (\ref{bra_eq}) may be re-expressed in the form of a set of generalized linear eigenvalue equations, 
\begin{equation}
\langle\Phi|\tilde{S}(e^{-S}He^{S}-E)C^{+}_{I}|\Phi\rangle\,, \quad \forall I \neq 0\,,  \label{bra_eq_alternative}
\end{equation}
for the
coefficients $\{\tilde{\cal S}_{I}\}$.

The GS energy eigenvalue $E$, is then just the value of $\bar{H}$ at
the minimum determined from solving equations (\ref{ket_eq}) and
(\ref{bra_eq}), namely
\begin{equation}
E=\langle\Phi|e^{-S}He^{S}|\Phi\rangle=\langle\Phi|He^{S}|\Phi\rangle\,.   \label{eq_E}
\end{equation}
Similarly, one may find the GS expectation value of any other operator
in terms of the coefficients $\{{\cal S}_{I},\tilde{\cal S}_{I}\}$.
For example, we may calculate the GS magnetic order parameter $M$,
which is defined to be the average local on-site magnetization,
\begin{equation}
M \equiv=-\frac{1}{N}\langle\Phi|\tilde{S}e^{-S}\sum^{N}_{l=1}s^{z}_{l}e^{S}|\Phi\rangle\,,  \label{M_eq}
\end{equation}
where in equation (\ref{M_eq}) the spins are referred to their local
rotated frames.

We note that no approximations have yet been made.  However, equations
(\ref{ket_eq}) for the coefficients $\{{\cal S}_{I}\}$ are explicitly
nonlinear, and one may wonder if truncations of the exponential terms
are needed in practice.  We note that these appear, in both equations
(\ref{ket_eq}) and (\ref{bra_eq_alternative}), only in the form of the
similarity transformation $e^{-S}He^{S}$ of the Hamiltonian.  This
combination may be expanded in terms of the well-known sum of nested
commutators.  Another key feature of the CCM is that this otherwise
infinite sum will now terminate exactly with the double commutator
term.  This is due to the basic SU(2) spin commutation relations,
together with the fact that all of the terms in equation
(\ref{correlation_oper}) that comprise the decomposition for the
operator $S$ both commute with one another and are simple products of
single-spin spin-raising operators.  Thus, all terms in the expansion
for $e^{-S}He^{S}$ are linked, and the Goldstone linked-cluster
theorem is exactly preserved even when the expansion of equation
(\ref{correlation_oper}) for $S$ is truncated in any conceivable way.
In turn, this guarantees that the CCM is size-extensive at any such
level of truncation, so that we may work in the thermodynamic limit
($N \rightarrow \infty$) from the very beginning.  Similar
considerations also apply to the evaluation of the GS expectation
value of any operator, such as that for the magnetic order parameter
$M$ in equation (\ref{M_eq}).

In practice, therefore, the sole approximation made is to restrict the
set of indices $\{I\}$ retained in the expansions of equation
(\ref{correlation_oper}) for the CCM correlation operators $\{S, \tilde{S}\}$.  As in our earlier work
\cite{Bishop:2012_checkerboard} on this model, and in many other
applications too, we use here the well-tested localized
(lattice-animal-based subsystem) LSUB$m$ truncation scheme
\cite{Bishop:2012_checkerboard,Bishop:1991_XXZ_PRB44,Zeng:1998_SqLatt_TrianLatt,SEKruger:2006_spinStiff,Darradi:2008_J1J2mod,Bi:2009_SqTriangle,Richter2010:J1J2mod_FM,Gotze:2011_kagome,Bishop:2013_crossStripe,DJJFarnell:2014_archimedeanLatt,Richter:2015_ccm_J1J2sq_spinGap},
in which at the $m$th level of approximation we retain only the
multispin-flip configurations $\{I\}$ in equation
(\ref{correlation_oper}) that are defined by at most $m$ contiguous
lattice sites.  A cluster configuration is defined to be contiguous in
this sense if every site in the cluster is adjacent (or connected) in
the NN sense (in the selected geometry) to at least one other site in
the cluster.

The number of such fundamental LSUB$m$ configurations,
$N_{f}=N_{f}(m)$, may be reduced by incorporating the space- and
point-group symmetries of the Hamiltonian and model state being used,
as well as any conservation laws that similarly pertain to both.
Nevertheless, the number $N_{f}(m)$ is a rapidly increasing function of the
truncation index $m$, and hence the need soon arises to utilize
massive parallelization plus supercomputing resources for the
highest-order calculations we undertake
\cite{Zeng:1998_SqLatt_TrianLatt,ccm_code}.  For the present
checkerboard model we employ both the N\'{e}el and N\'{e}el$^{\ast}$
states shown in figures \ref{model_bonds}(b) and \ref{model_bonds}(c),
respectively, as our CCM model states.  We also use the basic
checkerboard geometry to define the LSUB$m$ configurations, in the
sense that we treat both the pairs of sites connected by $J_{1}$ bonds
as well as those connected by $J_{2}$ bonds as being contiguous sites.
For both model states we are able to perform LSUB$m$ calculations with
values of the truncation index $m \leq 10$.

It is worth noting that if we were to use the basic square-lattice
geometry (i.e., with LSUB$m$ contiguous sites defined only by those
joined by $J_{1}$ bonds), the number $N_{f}(m)$ of fundamental
configurations would be considerably smaller than in the checkerboard
geometry, at the same LSUB$m$ level.  In turn, this could possibly
permit us to perform higher-order LSUB$m$ calculations with the same
level of computational resource.  However, this possible advantage is
accompanied by the severe drawback that the LSUB$m$ sequences of
approximations for both $E/N$ and $M$ now display a marked staggering
effect in $m \equiv 2k$, depending on whether the index $k$ is now odd
or even.  The reason for such a staggering behaviour is clearly due to
the fact that the full LSUB$m$ sequence using the square-lattice
geometry does not properly reflect the checkerboard-lattice
symmetries.  Such odd-even staggering effects in index $m$ have been
observed previously in CCM LSUB$m$ approximations for simple
(dynamically unfrustrated) models \cite{Farnell:2008_oddEvenOrd}.  It
always has the undesirable consequence of making extrapolations (for
the full sequence) to the exact $m \rightarrow \infty$ limit much more
difficult and much less robust.

Thus, finally, the only remaining step (and the only approximation
made in the whole CCM procedure) is to extrapolate our LSUB$m$
sequences of approximants for a given GS parameter to the limit $m
\rightarrow \infty$ where the method is exact.  For the GS energy we
use the usual, well-tested extrapolation scheme
\cite{Bishop:2012_checkerboard,Fa:2004_QM-coll,Darradi:2008_J1J2mod,Bi:2009_SqTriangle,Richter2010:J1J2mod_FM,Gotze:2011_kagome,Bishop:2013_crossStripe,DJJFarnell:2014_archimedeanLatt,Richter:2015_ccm_J1J2sq_spinGap,Farnell:2008_oddEvenOrd},
\begin{equation}
E(m)/N=a_{0}+a_{1}m^{-2}+a_{2}m^{-4}\,.           \label{E_extrapo}
\end{equation}
For highly frustrated magnetic systems, particularly ones which
are close to a quantum critical point (QCP) or for which the magnetic
order parameter $M$ is close to zero, the extrapolation scheme which
has been found to be appropriate in many earlier studies (see, e.g.,
\cite{Bishop:2012_checkerboard,Darradi:2008_J1J2mod,Richter2010:J1J2mod_FM,Gotze:2011_kagome,Bishop:2013_crossStripe,DJJFarnell:2014_archimedeanLatt,Richter:2015_ccm_J1J2sq_spinGap})
is one with a leading exponent $1/m^{1/2}$,
\begin{equation}
M(m)=c_{0}+c_{1}m^{-1/2}+c_{2}m^{-3/2}\,.            \label{M_extrapo_frustrated}
\end{equation}

We note finally that, without making any truncation (e.g., by using
the LSUB$m$ scheme utilized here), the solutions to equations
(\ref{ket_eq}) and (\ref{bra_eq_alternative}) are formally exact, and
hence independent, in principle, of the reference state $|\Phi\rangle$
used (at least as long, for example, as $|\Phi\rangle$ shares the same
quantum numbers as the exact state $|\Psi\rangle$).  The only caveat
is that this assumes intrinsically that a solution (to the full,
untruncated CCM equations) exists for a given choice of model state
$|\Phi\rangle$.  In practice, of course, approximations must be made,
e.g., by using the LSUB$m$ scheme advocated above.  Then, of course,
the solutions at a given LSUB$m$ level of truncation for
$|\Psi\rangle$ may well, in principle, depend on the reference state
$|\Phi\rangle$.  As discussed more fully below in section
\ref{results}, the range of values of the Hamiltonian parameters for
which the truncated equations have a solution usually does, in
practice, depend both on the choice of $|\Phi\rangle$ and the order
$m$ of the truncation.  So long as the scaling procedure to the $m
\rightarrow \infty$ limit would be exact, the dependence on
$|\Phi\rangle$ would then be removed.  In practice, since the scaling
laws are empirical, the best we can expect is that any remaining
dependence on any states $|\Phi\rangle$ for which solutions to the CCM
equations exist in the same parameter range will be only very weak.

\section{Results}
\label{results}
We now present our CCM results separately for the two cases when the model state is chosen to be the N\'{e}el state or the N\'{e}el$^{\ast}$ state.
\subsection{N\'{e}el model state; N\'{e}el phase boundaries}
In our previous paper on the spin-$\frac{1}{2}$ anisotropic planar
pyrochlore \cite{Bishop:2012_checkerboard} we set $J_{1}=+1$ to set
the energy scale for the case of AFM NN bonds, and employed the
N\'{e}el state as our CCM model state in order to investigate the QCP
at which N\'{e}el order vanishes as we turn on and strengthen NNN
bonds with a frustrating exchange coupling of strength $J_{2} \equiv
\kappa J_{1} > 0$.  In particular, we calculated the magnetic order
parameter $M$ at various LSUB$m$ levels of approximation for $m \leq
10$, for values of the frustration parameter $\kappa$ in the range $0
\leq \kappa < \kappa^{t}(m)$.  Here, $\kappa^{t}(m)$ is the highest
value for $\kappa$ for which, at a given LSUB$m$ level of
approximation, a real solution to the CCM equations (\ref{ket_eq}) could
be found.  Such termination points are by now well understood.  They
have been shown
\cite{Fa:2004_QM-coll,Bi:2009_SqTriangle,Richter2010:J1J2mod_FM} to be
direct manifestations of the associated QCP at which the order of the
model state melts in the physical system under study.  As is typically
the case, we found that the corresponding values of $\kappa^{t}({m})$
are greater than the associated critical value $\kappa_{c_{1}}$
($\approx 0.80 \pm 0.01$), and that they approach $\kappa_{c_{1}}$
monotonically (as a function of $m$) from above.

We now turn our attention first to the case when the NNN bonds become
FM in nature (i.e., $J_{2}<0$), still keeping $J_{1}=+1$.  Clearly,
the two bonds no longer frustrate one another and we fully expect the
N\'{e}el order to be preserved in such a way that the order parameter
$M$ monotonically increases as $|J_{2}|$ increases, with $M
\rightarrow \frac{1}{2}$ as $J_{2} \rightarrow -\infty$.  Then, as the
strength $J_{1}$ of the NN bonds changes sign in the same limit $J_{2}
\rightarrow -\infty$, the stable GS phase will clearly become the FM
phase via a first-order transition.  Since the FM states (of both the
entire lattice and the decoupled crossed 1D chains in this limit) are
exact eigenstates of the Hamiltonian of equation (\ref{H}) for all
values of the spin quantum number $s$, the phase boundary between the
N\'{e}el and FM phases is expected to be the same for the quantum case
as in figure \ref{model_bonds}(e) for the classical ($s \rightarrow
\infty$) case.

\begin{figure}[!t]
\mbox{
   \subfigure[]{\scalebox{0.3}{\includegraphics[angle=270]{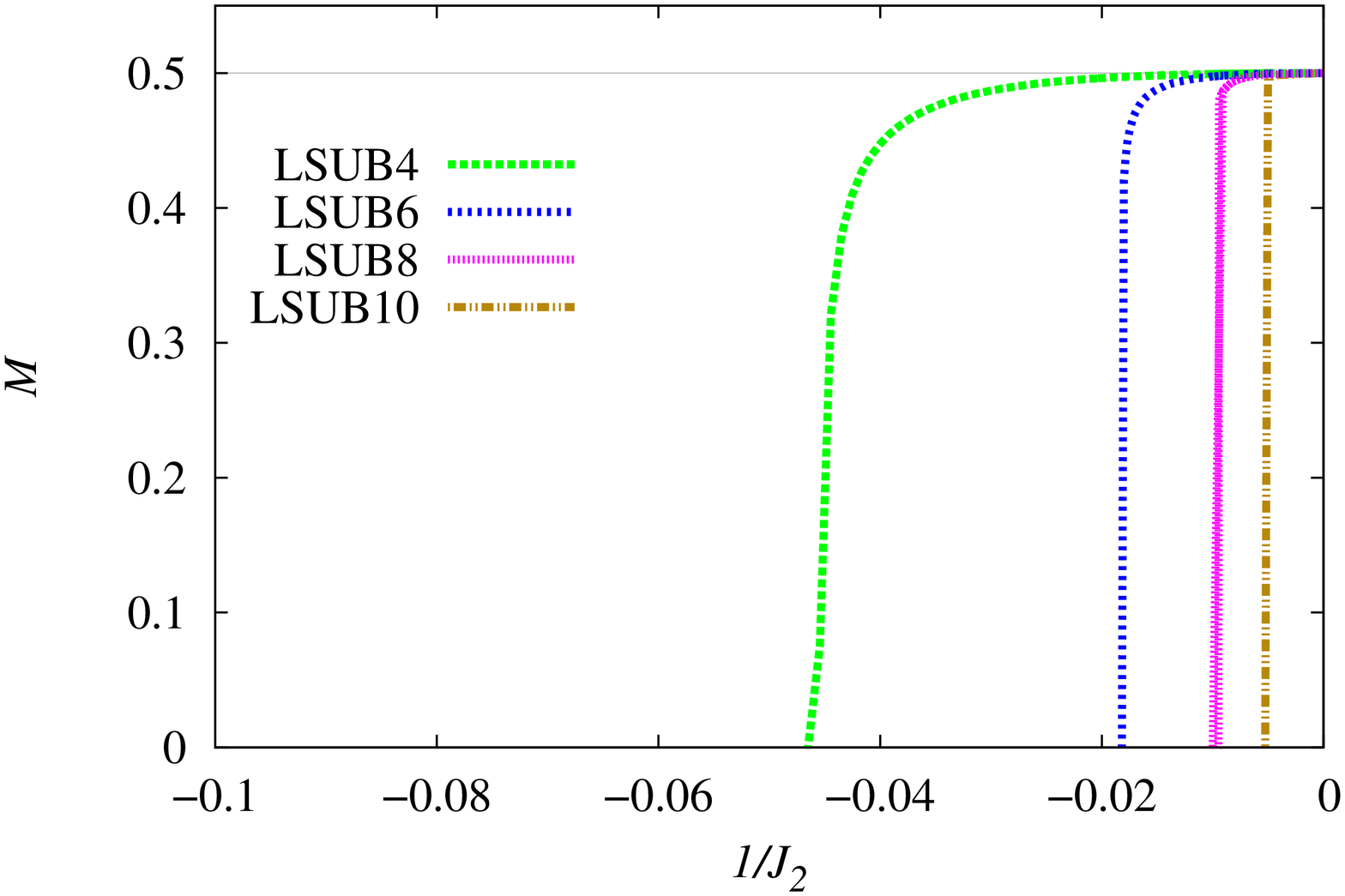}}}
\subfigure[]{\scalebox{0.3}{\includegraphics[angle=270]{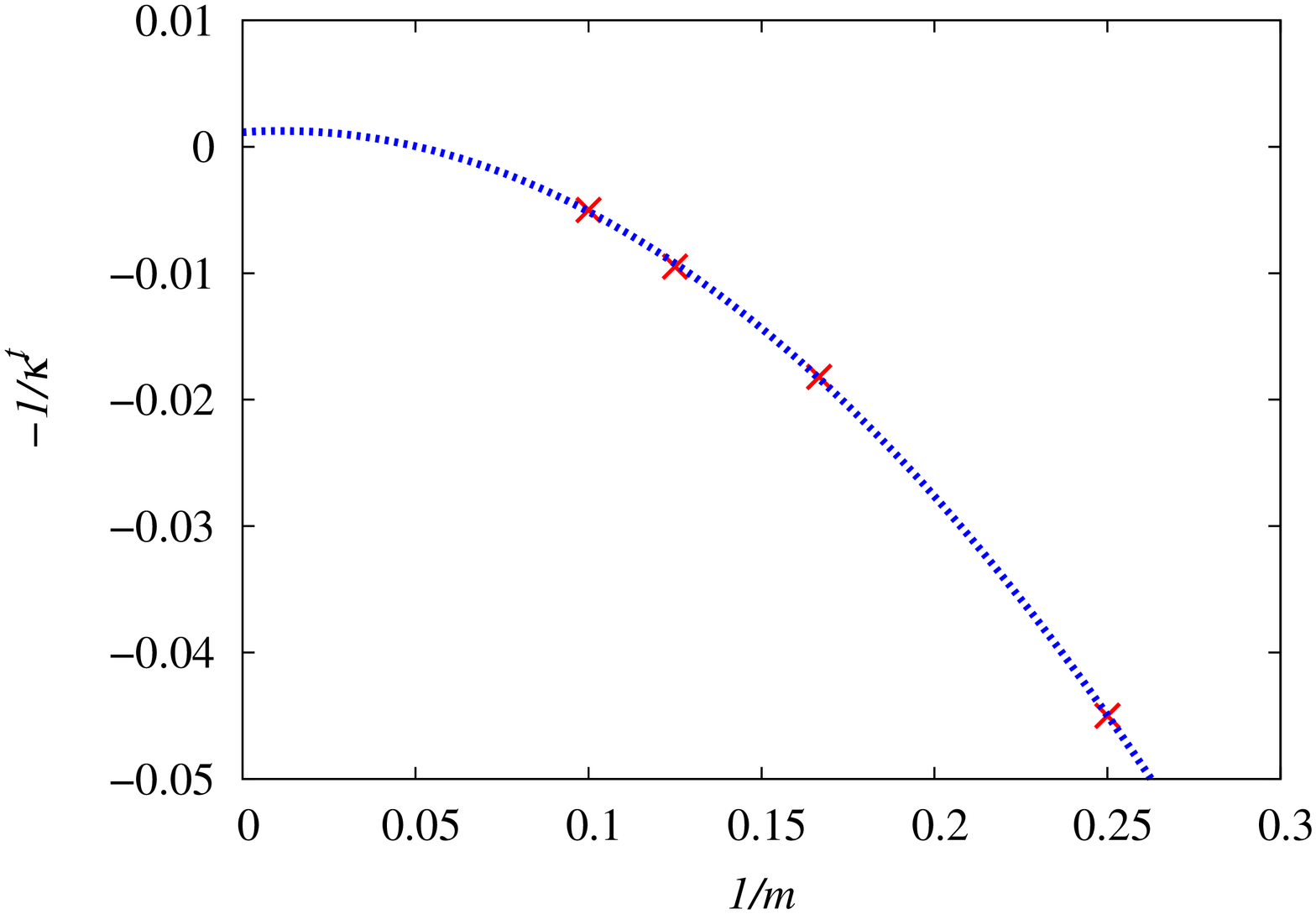}}}
}
\caption{(a) Ground-state magnetic order parameter, $M$, 
   for the N\'{e}el state of the spin-$\frac{1}{2}$ $J_{1}$--$J_{2}$ checkerboard model
   (with $J_{1}=
-1$, $J_{2}<0$) as a function of $1/J_{2} \equiv -1/\kappa$.  The CCM results for
   various LSUB$m$ approximations with $m=4,6,8,10$ are shown.  (b) The scaling of the LSUB$m$ termination points $\kappa^{t}(m)$ in figure \ref{M_J2neg}(a) as a function of $1/m$, is shown according to equation (\ref{kappa_t-scale}).}
\label{M_J2neg}
\end{figure}

Nevertheless, it is of interest to see how our CCM LSUB$m$ solutions
based on the N\'{e}el state as model state actually conform to the
above expectation.  Accordingly, we show in figure \ref{M_J2neg}(a) the
LSUB$m$ results for the magnetic order parameter $M$, for the cases
$m=\{4,6,8,10\}$, with $J_{1}=-1,\,J_{2}<0$.
We see once again very clearly that the LSUB$m$ results based on the
N\'{e}el state terminate slightly into the FM regime, with the
corresponding lower termination points monotonically approaching the
N\'{e}el-FM boundary at $\kappa_{c_{4}} = \infty$ as $m \rightarrow
\infty$.  The actual values of the inverse of the lower LSUB$m$
termination points, $1/\kappa^{t}(m)$, are +0.045, +0.0182, +0.00948
and +0.00500 for $m=4,6,8$, and 10, respectively.  Clearly, any
reasonable extrapolation scheme is compatible with
$1/\kappa^{t}(\infty)=0$.  For example, the scheme
\begin{equation}
1/\kappa^{t}(m)=\alpha_{0}+\alpha_{1}m^{-1}+\alpha_{2}m^{-2} \,, \label{kappa_t-scale}
\end{equation}
shown in figure \ref{M_J2neg}(b), yields
the value $\alpha_{0}=-0.0011 \pm 0.0019$, where the error is simply
that associated with the fit to the four values $m=\{4,6,8,10\}$.

The fact that we find solutions of the CCM LSUB$m$ equations based on
the N\'{e}el state as reference state (with a non-vanishing value of
the N\'{e}el order parameter) for small negative values of $J_{2}$
(i.e., intruding slightly into the FM regime), is related to the fact
that our solutions are thus intrinsically biased towards the stability
of the N\'{e}el phase.  As one proceeds to higher LSUB$m$ orders
(i.e., as the index $m$ is increased) the region of unphysical
intrusion decreases, and vanishes in the $m \rightarrow \infty$ limit,
as may clearly be observed in figure \ref{M_J2neg}(a).  This is a
completely general feature of CCM LSUB$m$ solutions, which has been
observed in many previous CCM studies of quantum magnets.  In general,
if the stable GS phase (say, {\it B}) in the physical intrusion region
based on a given CCM reference state (say, {\it A}) is itself amenable
to a CCM solution based on another (typically classical) reference
state {\it B}, then we would also typically find a similar region of
unphysical intrusion of the latter {\it B}-phase solution into the
physical {\it A}-phase region.  However, in the present case, since
the FM wave function illustrated in figure \ref{model_bonds}(a) is
itself an exact eigenstate of our Hamiltonian of equation (\ref{H}), it
cannot be used as a CCM reference state, except in the trivial sense
that all (exact or approximate) solutions based on it simply yield
vanishing values of all of the correlation coefficients $\{{\cal
  S}_{I},{\tilde{\cal S}}_{I}\}$ retained in equation
(\ref{correlation_oper}).

\subsection{N\'{e}el$\,^{\ast}$ model state; CDVBC phase boundaries}
We now turn to our corresponding LSUB$m$ results based on the
N\'{e}el$^{\ast}$ state of figure \ref{model_bonds}(c) as the CCM
model state.  In our previous paper \cite{Bishop:2012_checkerboard} we
investigated the case of AFM NN bonds ($J_{1}=+1$), and we now
similarly consider the case of FM NN bonds ($J_{1}=-1$).  In both
cases we investigate the effect of frustrating AFM NNN bonds
($J_{2}>0$).

\begin{figure}[!t]
\mbox{
   \subfigure[]{\scalebox{0.3}{\includegraphics[angle=270]{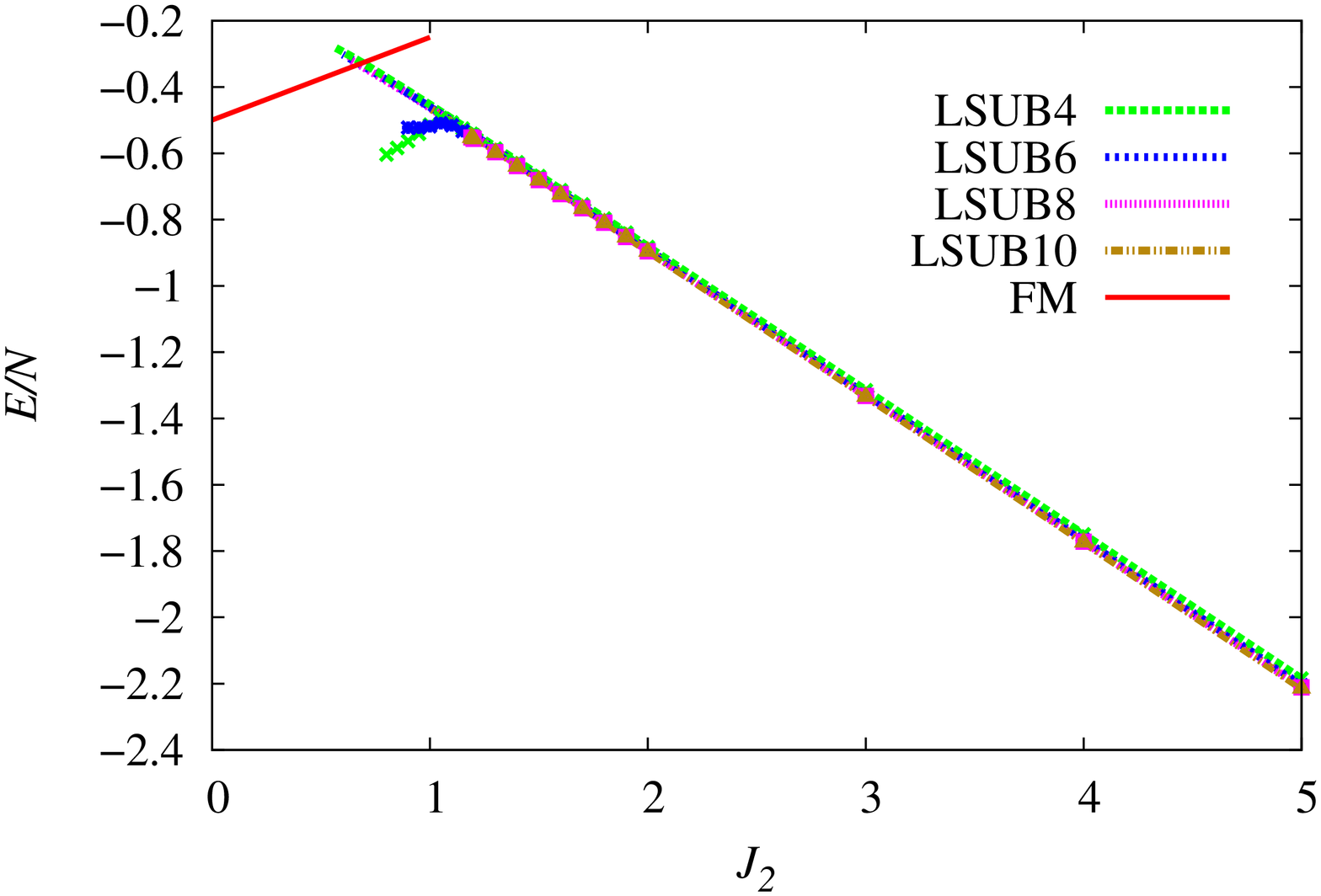}}}
\subfigure[]{\scalebox{0.3}{\includegraphics[angle=270]{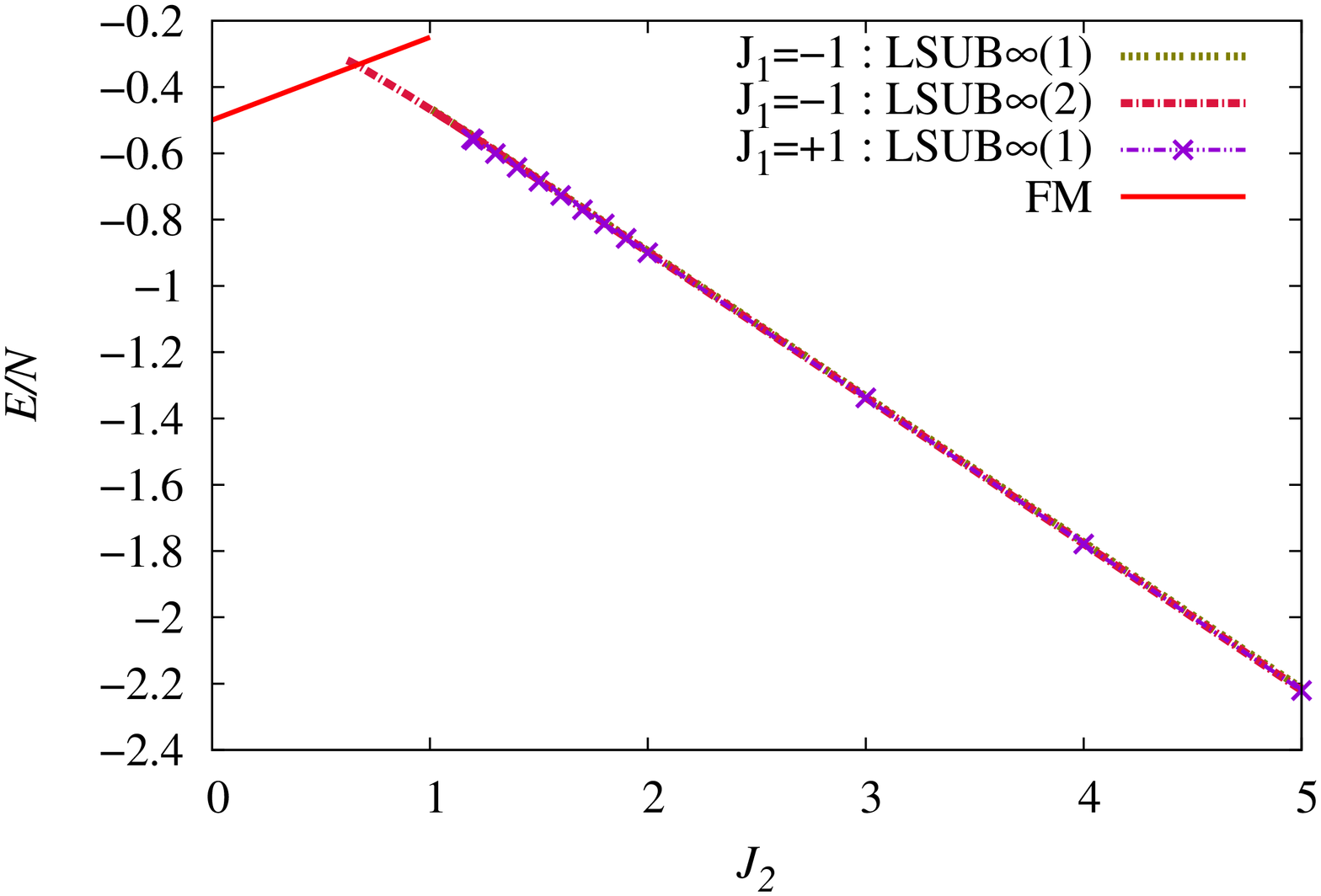}}}
}
\caption{CCM results for the GS energy per spin, $E/N$, for the spin-$\frac{1}{2}$ $J_{1}$--$J_{2}$
  Heisenberg model on the checkerboard lattice.   (a) LSUB$m$ results with $m=4,6,8,10$ based on the
  N\'{e}el$^{\ast}$ state.  Results without symbols attached refer
  to the case $J_{1}=-1$, and are compared to the corresponding results for the case $J_{1}=+1$, shown with symbols attached.  The GS energy per spin, $E/N=(J_{2}-2)/4$, for the FM phase is also shown.  (b) The extrapolated results using equation (\ref{E_extrapo}): LSUB$\infty$(1) uses the
   data sets $m=\{4,6,8,10\}$ and LSUB$\infty$(2) uses the
   data sets $m=\{2,4,6,8\}$.}
\label{E}
\end{figure}

Results for the ground-state energy per spin, $E/N$, for both cases
($J_{1}=-1$ and $J_{1}=+1$), are shown in figure \ref{E}.
Figure \ref{E}(a) shows the ``raw'' LSUB$m$ results for values of the
truncation index $m=4,6,8,10$, while figure \ref{E}(b) shows
extrapolated LSUB$\infty$ results using equation (\ref{E_extrapo}).
We also show the exact result for the FM phase for the present
$s=\frac{1}{2}$ model, $E_{{\rm FM}}/N=\frac{1}{4}(2J_{1}+J_{2})$,
with $J_{1}=-1$.  In figure \ref{E} the curves without symbols
attached refer to our present case ($J_{1}=-1$), while the
corresponding curves with symbols attached refer to the case
$J_{1}=+1$.  They are taken from our earlier work
\cite{Bishop:2012_checkerboard}, and are shown for the sake of comparison.

Figure \ref{E}(a) clearly shows that the LSUB$m$ results for the GS
energy converge extremely rapidly in both cases as $m$ is increased.
Furthermore, we see that for all values $J_{2} \gtrsim 1.5$ the results for
both cases are remarkably similar.  Nevertheless, important
qualitative differences emerge for small values of $J_{2}$, in the
vicinity of the corresponding classical phase transitions at $J_{2}=1$
in both cases, as shown in figure \ref{model_bonds}(e).  In figure
\ref{E}(a) both sets of results are shown out to their respective
(approximately determined) termination points, $|\kappa^{t}(m)|$.  It
is evident that the value of $|\kappa^{t}(m)|$ for a given value of
$m$ is always less in the case $J_{1}=-1$ than in the case $J_{1}=+1$.
As we showed previously \cite{Bishop:2012_checkerboard}, in the case
$J_{1}=+1$, the values $\kappa^{t}(m) \rightarrow \kappa_{c_{2}}
\approx 1.22 \pm 0.02$ as $m \rightarrow \infty$.

By contrast, the respective LSUB$m$ values of $\kappa^{t}(m)$ based on
the N\'{e}el$^{\ast}$ state in the case $J_{1}=-1$ are $\kappa^{t}(4)
\approx -0.56$, $\kappa^{t}(6) \approx -0.60$, and $\kappa^{t}(8)
\approx -0.63$.  These termination points are themselves very close to
the corresponding values where the GS energy curves cross that of the
FM state, $\kappa^{e}(4) \approx -0.692$, $\kappa^{e}(6) \approx
-0.677$, and $\kappa^{e}(8) \approx -0.671$.  Once again, as the LSUB$m$
truncation index $m$ is increased to infinity, the size of the
region in which our solutions extend into the FM regime diminishes to
zero, and we expect that
$\kappa^{t}(\infty)=\kappa^{e}(\infty)=\kappa_{c_{3}}$.

We note, however, that it is computationally very costly to determine
the values $\kappa^{t}(m)$ with high accuracy, since the CCM LSUB$m$
solutions require increasingly greater amounts of computational
resource, for a specified level of accuracy, the closer a termination
point is approached.  For this reason the values $\kappa^{e}(m)$ are
appreciably more accurate than the corresponding values
$\kappa^{t}(m)$.  We also note that, for the same reason, we have been
unable to track the LSUB10 solution based on the N\'{e}el$^{\ast}$ state, in
the case $J_{1}=-1$, down to values sufficiently close to
$\kappa^{t}(10)$ for the respective energy curve to have crossed the
FM energy curve.
A simple extrapolation of the energy crossing points $\kappa^{e}(m)$,
using a scheme of the form of equation (\ref{E_extrapo}) and the
LSUB$m$ values with $m=\{4,6,8\}$ yields a value $\kappa_{c_{3}} \approx -0.661$ for the QCP.  A similar extrapolation using the
data set $m=\{3,5,7\}$ gives the value $\kappa_{c_{3}} \approx -0.662$.

While it is certainly true that three-parameter fits to only three
$m$-value points, as quoted above, are intrinsically dangerous, the
energy curves themselves are very smooth, even near the critical point
$\kappa_{c_{3}}$, as one can see clearly from figure \ref{E}.  Hence,
we may have considerable confidence in the quoted values.  However, to
verify this point we may also use the extrapolated curve based on the
N\'{e}el$^{\ast}$ state with the four-point data set $m=\{4,6,8,10\}$.
Since the calculated LSUB10 curve now terminates before the crossing
point with the FM energy curve, the resulting extrapolated
N\'{e}el$^{\ast}$ energy curve must itself be extrapolated the short
distance to the crossing point, which introduces an additional error.
Nevertheless, since the curve itself is again very smooth, and indeed
almost linear over the entire range of $J_{2}$ values shown in figure
\ref{E}(b), such an extrapolation (e.g., using simple polynomial fits)
is, in fact, rather robust (e.g., with respect to the order of the
fitting polynomial used).  The resulting values of $\kappa_{c_{3}}$
for fits using polynomials between third and sixth orders, for
example, all lie in the range -0.67 to -0.68.  They are thus
compatible with the (more accurate) values obtained above from the two
fits using three $m$-values, which require no such extrapolation to
the crossing point.  Such considerations, together with a more
detailed error analysis, leads us to our CCM estimate $\kappa_{c_{3}}
= -0.66(1)$.

We observe from figure \ref{E}(b) that the energy curves rapidly
approach the $J_{2} \rightarrow \infty$ limit of uncoupled
spin-$\frac{1}{2}$ 1D HAFM crossed chains, even for values of $J_{2}
\gtrsim 2$.  For example, the extrapolated value of the energy per
spin in the regime $4 \leq J_{2} \leq 5$ is $E/N \approx
-0.4420J_{2}$, which already is in excellent agreement with the exact
asymptotic result, $E/N=(\frac{1}{4}-\ln2)J_{2} \approx -0.4431J_{2}$
from the Bethe ansatz solution \cite{Bethe:1931,Hulthen:1938}.
Indeed, if we were to include this curve in figure \ref{E}(b) it would
lie virtually on top of the extrapolated LSUB$\infty$ curves, over the
whole of the range $1 < J_{2} < 5$, and deviating only very slightly
from them for values $J_{2} \lesssim 1$.

Although we have been able to use the N\'{e}el$^{\ast}$ state
completely successfully as a CCM model state, we now need to
investigate the stability of the quasiclassical magnetic LRO in the
regimes in which we have used it.  To that end we now show in figure
\ref{M}(a) the corresponding results for the GS magnetic
order parameter, $M$, to those shown in figure \ref{E}(a) for the GS
energy per spin, $E/N$.
\begin{figure*}[tb]
\begin{center}
\mbox{
  \subfigure[]{\scalebox{0.3}{\includegraphics[angle=270]{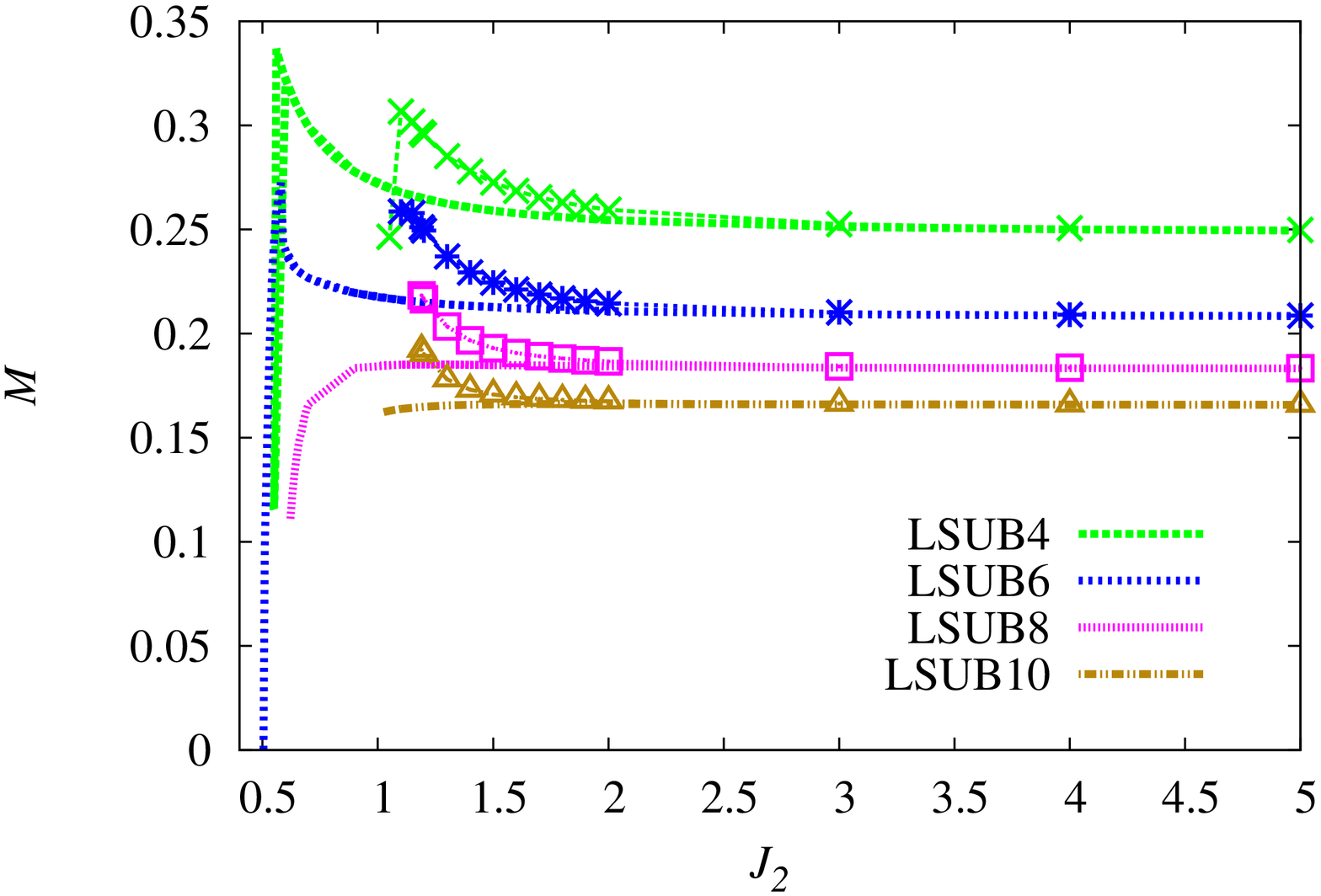}}}
\hfil
\subfigure[]{\scalebox{0.3}{\includegraphics[angle=270]{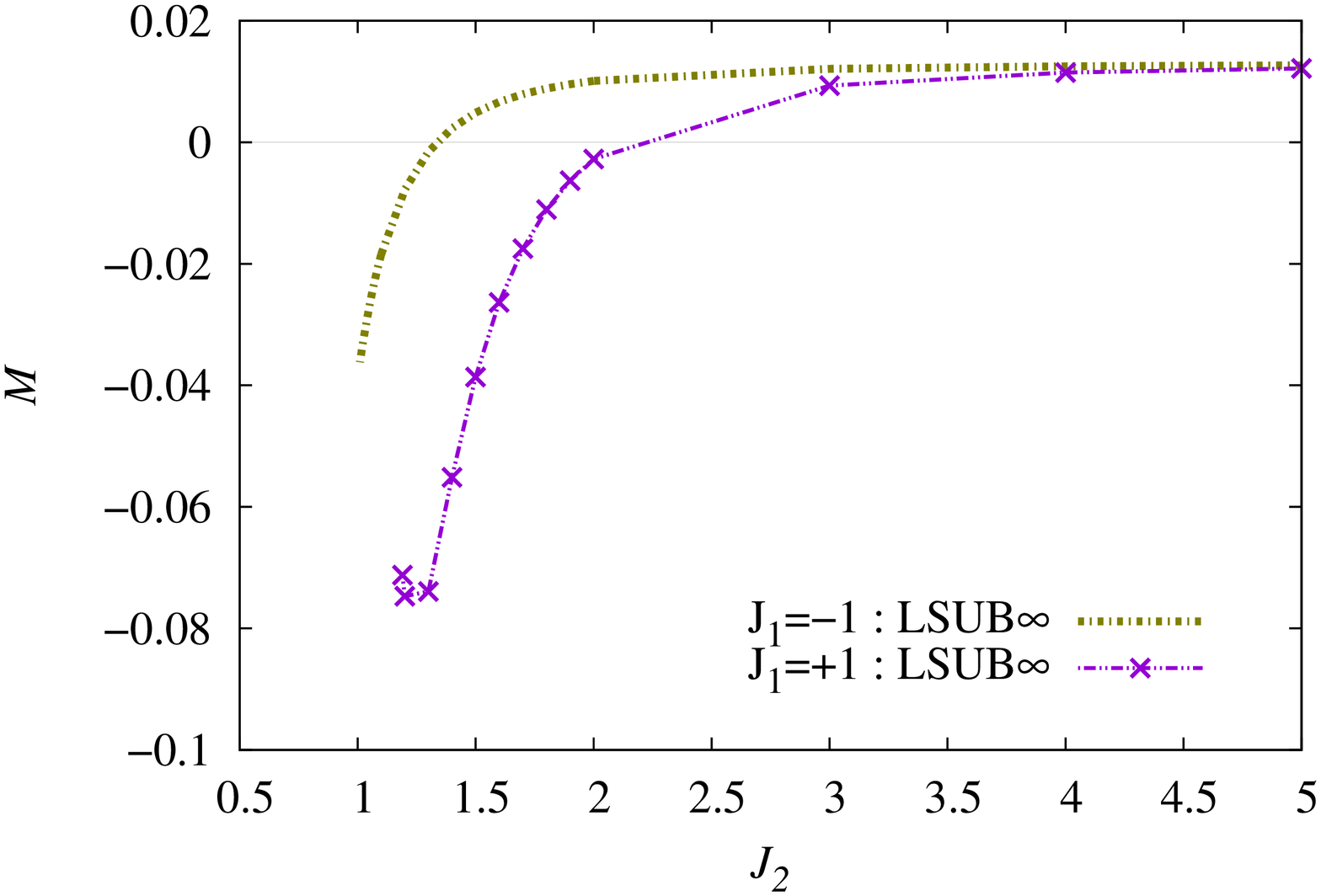}}}
}
 \caption{CCM results for the GS magnetic order parameter, $M$, 
  for the spin-$\frac{1}{2}$ $J_{1}$--$J_{2}$ Heisenberg model on the checkerboard lattice.  (a) LSUB$m$ results with $m=4,6,8,10$ based on the N\'{e}el$^{\ast}$ state.  Results without symbols attached refer to the case $J_{1}=-1$, and are compared to the corresponding results for the case $J_{1}=+1$, shown with symbols attached.  (b) The corresponding extrapolated results using equation (\ref{M_extrapo_frustrated}) and the data sets $m=\{4,6,8,10\}$.}
\label{M}
\end{center}
\end{figure*} 
It is clear that the CCM LSUB$m$ results for $M$ converge considerably
more slowly as a function of the truncation parameter $m$ than do
those for $E/N$, as is to be expected by a comparison of the
respective extrapolation schemes of equations
(\ref{M_extrapo_frustrated}) and (\ref{E_extrapo}).  Once again, just
like the previous results for $E/N$ in figure \ref{E}(a), the
corresponding results for $M$ in figure \ref{M}(a) are seen to be
remarkably similar for the two cases $J_{1}=-1$ and $J_{1}=+1$ in the
region $J_{2} \gtrsim 2$.  The divergence between the two cases for
values $J_{2} \lesssim 2$ is completely consistent with our previous
discussion for the GS energy results.

In our earlier work \cite{Bishop:2012_checkerboard} for the case
$J_{1}=+1$ we showed that the rapid rise and then precipitous fall in
each of the LSUB$m$ results for $M$, as $J_{2}$ approaches the
corresponding termination point for that solution, is a clear marker
of the associated QCP at $\kappa_{c_{2}} \approx 1.22$.  Similarly,
the analogous behaviour seen in figure \ref{M}(a) for the LSUB$m$
results for $M$ for the present case $J_{1}=-1$ is evidently
associated with the QCP at $\kappa_{c_{3}} \approx -0.66$ that we have
found above in the corresponding results for the GS energy.

In figure \ref{M}(b) we show the respective extrapolated
LSUB$\infty$ results for $M$ for the two cases, as obtained from the
scheme of equation (\ref{M_extrapo_frustrated}), used with the
respective LSUB$m$ data sets with $m=\{4,6,8,10\}$.  We see that $M$
is either zero (or very close to zero) or negative over the entire
range shown in figure \ref{M}, in both cases $J_{1}=-1$ and
$J_{1}=+1$.  Our results are, in particular, completely compatible
with $M$ being zero in both cases in the asymptotic limit $J_{2}
\rightarrow \infty$.  This is precisely the exact result for this
asymptotic (Luttinger spin-liquid) limit of decoupled 1D HAFM chains.
We thus conclude that the N\'{e}el$^{\ast}$ state is unlikely to be
the stable GS phase for any values of the parameters $J_{1}$ and
$J_{2}$ for which we have, nevertheless, successfully been able to use
it as a CCM model state.

Using techniques that combine renormalization group ideas, plus 1D
bosonization and current algebra methods, together with a careful
analysis of the relevant terms near the delicate Luttinger liquid
fixed point of the 1D HAFM spin chain, Starykh {\it et al}. \cite
{Starykh:2005_chkboard} showed that in the large-$\kappa$ limit the
stable GS phase exhibits spontaneous dimerization.  The GS phase in
this limit has CDVBC order with twofold spontaneous symmetry breaking,
which comprises a staggered ordering of dimers along the crossed
$J_{2}$ chains (i.e., along the diagonals in figure
\ref{model_bonds}).  This finding of a CDVBC state with no magnetic
order was then independently confirmed
\cite{Arlego:2007_chkboard,Arlego:2009_chkboard} in an analysis using
a SE technique based on the flow equation method.

It is thus of interest to investigate within our present CCM analysis
whether the whole (or part) of the regime that has been accessible to
us using the N\'{e}el$^{\ast}$ state as model state, but for which we
have shown has no N\'{e}el$^{\ast}$ magnetic LRO, might in fact have
CDVBC order instead.  To that end we consider the response of the
system when a field operator, $F = \delta \hat{O}_{d}$, is added as a
small perturbation to the original Hamiltonian of equation (\ref{H}),
with $\delta$ an infinitesimally small $c$-number, and the operator
$\hat{O}_{d}$, illustrated in figure \ref{X_d}(b), promotes the
formation of CDVBC order.
\begin{figure}[!t]
\begin{center}
\mbox{
\subfigure[]{\scalebox{0.31}{\includegraphics[angle=270]{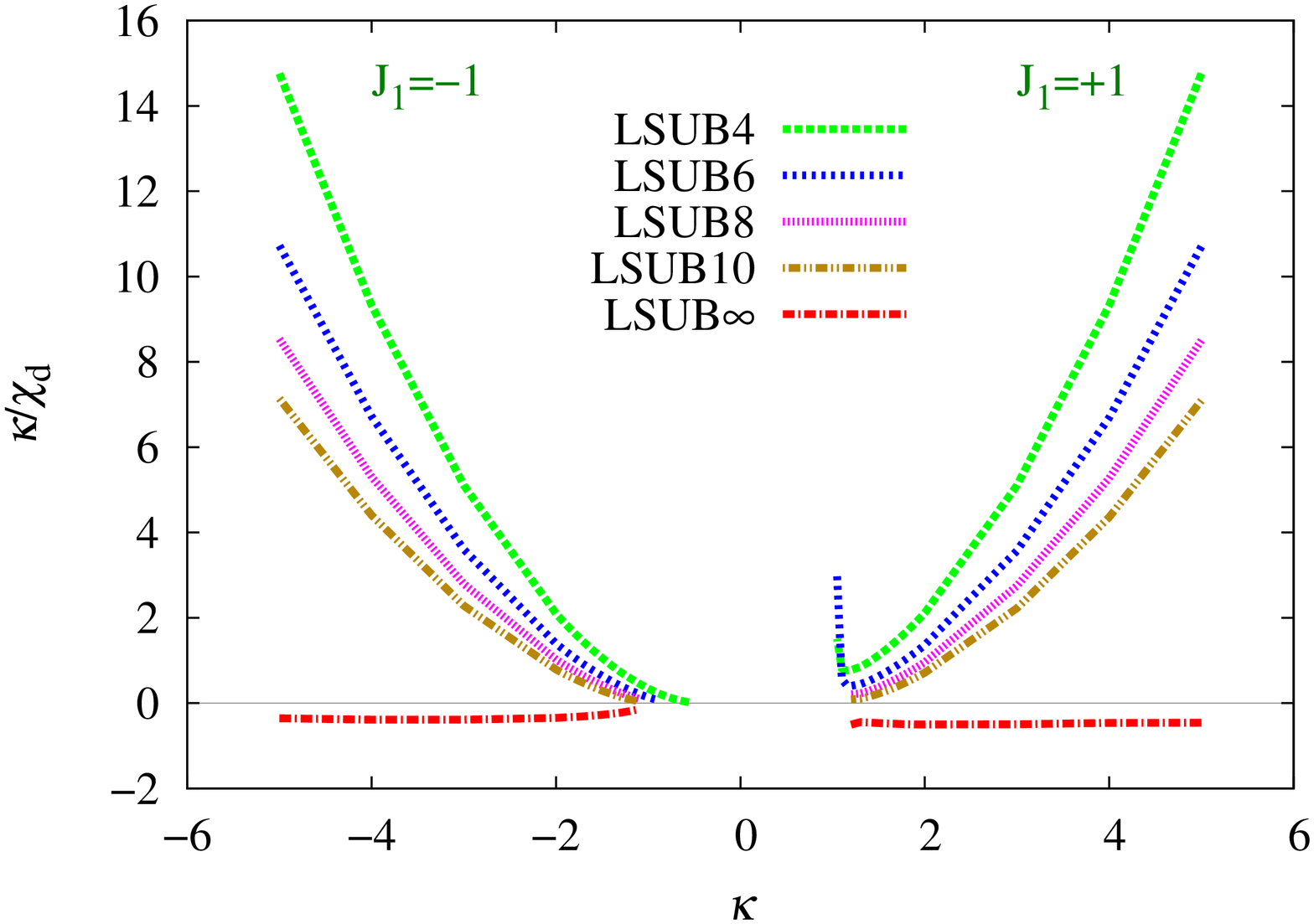}}}
\raisebox{-3.5cm}{
\subfigure[]{\scalebox{0.31}{\includegraphics{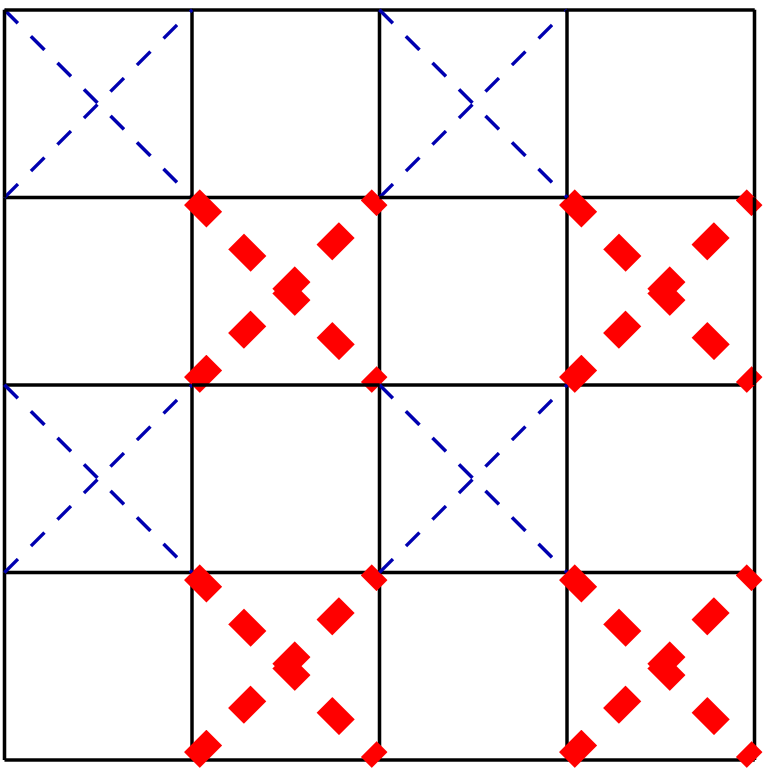}}}
}
}
\caption{(a) CCM results for the scaled inverse crossed-dimer
  susceptibility, $\kappa/\chi_d$, as a function of the frustration parameter, $\kappa \equiv J_{2}/J_{1}$, using the N\'{e}el$^{\ast}$ state as the CCM model 
  state, for the spin-$\frac{1}{2}$ $J_{1}$--$J_{2}$ Heisenberg
  model on the checkerboard lattice with $J_{1}=-1$ (left curves), comparing with those for $J_{1}=+1$ (right curves) (with $J_{2}>0$
   for both cases).  We also show the extrapolated LSUB$\infty$ results based on equation (\ref{Extrapo_inv-chi-2}) with the data sets $m=\{4,6,8,10\}$ as input.  (b) The perturbations (fields) $F=\delta\, \hat{O}_d$ for 
  the dimer susceptibility $\chi_d$.  Thick (red) dashed
  and thin (blue) dashed lines correspond respectively
  to strengthened and weakened NNN exchange couplings, where $\hat{O}_{d} =
  \sum_{\langle \langle i,k \rangle \rangle} a_{ik}
  \mathbf{s}_{i}\cdot\mathbf{s}_{k}$, and the sum runs over the NNN diagonal
  bonds of the checkerboard lattice, with $a_{ik}=+1$ and $-1$ for thick (red) 
  dashed and thin (blue) dashed lines respectively.  The
  original solid (black) $J_{1}$ bonds are unaltered in strength.}
\label{X_d}
\end{center}
\end{figure}    

Thus, we now calculate the perturbed GS energy per spin, $e(\delta)
\equiv E(\delta)/N$, for the perturbed Hamiltonian, $H(\delta) \equiv
H + F$, at various LSUB$m$ levels of approximation, using the
N\'{e}el$^{\ast}$ state as our model state.  We then calculate the
corresponding susceptibility coefficient,
\begin{equation}
\chi_{d} \equiv -\left.\frac{\partial^{2}e(\delta)}{\partial\delta^{2}}\right|_{\delta=0}\,,  \label{Eq_X}
\end{equation}
which is a measure of the susceptibility of the system against the
formation of CDVBC order.  Clearly, $\chi_{d}$ can be used to find
points or regions in phase space where the phase corresponding to the
particular CCM model state used (viz., here the N\'{e}el$^{\ast}$
state) becomes unstable with respect to a CDVBC-ordered state, namely
when the extrapolated inverse susceptibility, $1/\chi_{d}$, goes to
zero.

In order to extrapolate our LSUB$m$ results to the LSUB$\infty$ limit,
it has been found \cite{DJJF:2011_honeycomb,Li:2013_chevron} that
the simple scheme,
\begin{equation}
\chi^{-1}_{d}(m) = x_{0}+x_{1}m^{-2}+x_{2}m^{-4}\,,  \label{Extrapo_X_asE}
\end{equation}
gives excellent fits, except in regions where $\chi^{-1}_{d}$ becomes
very small or zero.  In the present case we are, of course, interested
in precisely such regimes, and it is then preferable
\cite{DJJF:2011_honeycomb,Li:2013_chevron,Bishop:2013_crossStripe} to
use an unbiased extrapolation scheme of the form,
\begin{equation}
\chi^{-1}_{d}(m) = y_{0}+y_{1}m^{-\nu}\,,            \label{Extrapo_inv-chi-2}
\end{equation}
where the leading exponent $\nu$ is itself a fitting parameter, together with the coefficients 
$y_{0}$ and $y_{1}$.

We show our results for $\chi^{-1}_{d}$ in figure \ref{X_d}(a).  Since
the GS energy scales linearly with $J_{2}$ in the large $J_{2}$ limit,
as seen from figure \ref{E}, it is more appropriate to show our CCM
results in figure \ref{X_d}(a) for the scaled inverse dimer
susceptibility, $\kappa/\chi_{d}$, as a function of $\kappa$.  Figure
\ref{X_d}(a) shows results for both the cases $J_{1}=-1$ and
$J_{1}=+1$.  It is evident from LSUB$\infty$ extrapolations using
equation (\ref{Extrapo_inv-chi-2}) that our results in both cases are
completely consistent with $1/\chi_{d}$ being zero for all values of
$\kappa$ shown.  In other words, our results strongly indicate that
everywhere we have been able to use the N\'{e}el$^{\ast}$ state as a
CCM model state it is actually unstable against the formation of CDVBC
order, which forms the ordering of the true stable GS phase in these
regimes.  Hence, we conclude that at $\kappa_{c_{2}} = 1.22 \pm 0.02$
  there is a QCP between states with PVBC and CDVBC forms of order,
  and at $\kappa_{c_{3}} = -0.66 \pm 0.01$ there is a QCP between
  states with CDVBC and FM forms of order.

\section{Summary and conclusions}
\label{summary}
In this and our previous paper \cite{Bishop:2012_checkerboard} we have
used the CCM, implemented to high orders, to study the
spin-$\frac{1}{2}$ $J_{1}$--$J_{2}$ Heisenberg model on the
checkerboard lattice.  Whereas in our earlier work, we studied the
model only in the case where both NN and NNN bonds were AFM in nature
($J_{1}>0, J_{2}>0$), we have now investigated the model over the
entire $J_{1}$--$J_{2}$ phase space.  Previously we showed that the
classical phase transition at $\kappa^{a}_{{\rm cl}} = 1$ ($J_{1}>0$)
between the N\'{e}el AFM phase and the IDF of AFM phases, is split in
the $s=\frac{1}{2}$ model into two transitions at $\kappa_{c_{1}}=0.80
\pm 0.01$ and $\kappa_{c_{2}}=1.22 \pm 0.02$.  For $\kappa <
\kappa_{c_{1}}$ the N\'{e}el order persists, while for $\kappa_{c_{1}}
< \kappa < \kappa_{c_{2}}$ the GS phase has PVBC order.  Finally, we
showed that for $\kappa > \kappa_{c_{2}}$ (with $J_{1}>0$) the stable
GS phase has CDVBC order.  The transitions at both QCPs
$\kappa_{c_{1}}$ and $\kappa_{c_{2}}$ are likely to be direct ones,
although we cannot exclude very narrow coexistence regions confined to
$0.79 \lesssim \kappa \lesssim 0.81$ and $1.20 \lesssim \kappa
\lesssim 1.22$, respectively.

We have now confirmed that the CDVBC phase persists when the $J_{1}$
bonds become FM in nature ($J_{1}=-1$) to a QCP at
$\kappa_{c_{3}}=-0.66 \pm 0.01$, at which point the CDVBC phase gives
way to the FM phase.  This QCP may be compared with the corresponding
transition in the classical ($s \rightarrow \infty$) model at
$\kappa^{b}_{{\rm cl}}=-1$ between the IDF of AFM states and the FM
state.  It is interesting to note that quantum fluctuations act to
destabilize the FM ordering at a weaker level of frustration than for
the classical version of the model.  Precisely the same effect has now
also been seen in a variety of other comparable models, such as the
spin-$\frac{1}{2}$ $J_{1}$--$J_{2}$ model on the square lattice
\cite{Richter2010:J1J2mod_FM}, the spin-$\frac{1}{2}$
$J_{1}$--$J_{2}$--$J_{3}$ model (with $J_{3} = J_{2}$) on the
honeycomb lattice \cite{Bishop:2012_honey_phase}, and the
spin-$\frac{1}{2}$ $J_{1}$--$J_{2}$ model on a cross-striped square
lattice \cite{Bishop:2013_crossStripe}.  Finally, in the unfrustrated
region where $J_{2}<0$, the QCP between the FM and N\'{e}el phases has
been shown to occur at $\kappa_{c_{4}}=\infty$, at exactly the same
place as in the classical model ($\kappa^{c}_{{\rm cl}}$), fully as
expected.  Our results are summarized in the complete GS ($T=0$) phase
diagram shown in figure \ref{phase-diagram}.
\begin{figure}[!t]
\begin{center}
\includegraphics[width=12cm]{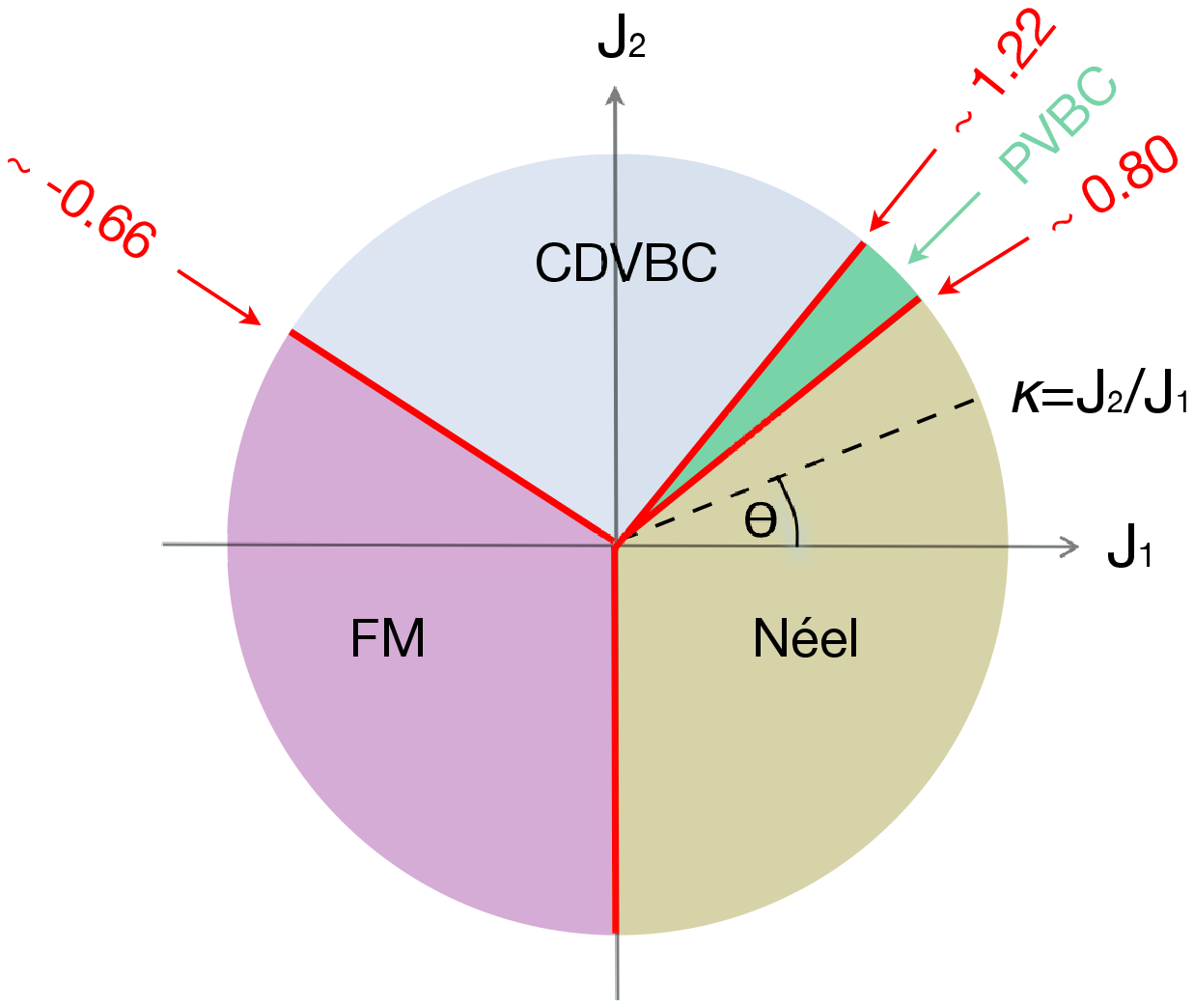}
\caption{The ground-state ($T=0$) phase diagram of the
  spin-$\frac{1}{2}$ $J_{1}$--$J_{2}$ Heisenberg model on a
  checkerboard lattice (with $\kappa \equiv J_{2}/J_{1} \equiv
  \tan\theta$), showing the N\'{e}el antiferromagnetic phase, the
  plaquette valence-bond crystalline (PVBC) phase, the crossed-dimer
  valence-bond crystalline (CDVBC) phase, and the ferromagnetic (FM)
  phase.  All of the transitions at $\kappa_{c_{1}} \approx 0.80(2)$,
  $\kappa_{c_{2}} \approx 1.22(2)$, and $\kappa_{c_{3}} \approx
  -0.66(1)$, each with $J_{2}>0$, appear to be direct ones.  The
  direct first-order transition between the FM and N\'{e}el phases is
  exactly at $\theta_{c_{4}}=\frac{3}{2}\pi$.}
\label{phase-diagram}
\end{center}
\end{figure}

We note that there have been suggestions in the literature
\cite{Shannon:2006_FM,Sindzingre:2007_FM,Sindzingre_Shannon:2009_FM,Sindzingre:2009_FM,Shindou_Momoi:2011_FM}
that the competition between FM interactions between NN pairs of spins
($J_{1}>0$) and AFM interactions between other pairs of spins in
frustrated spin-$\frac{1}{2}$ systems on the square lattice might
result in gapless spin-liquid states with multipolar order (e.g.,
spin-nematic states) near to the FM boundary.  Similar states have
also been hypothesised in frustrated multiple cyclic spin-exchange
models on the triangular lattice with FM NN pairwise interactions
\cite{Shindou_Momoi:2011_FM}, either in a nonzero external magnetic
field (with octupolar ordering occurring) or in zero field (with
quadratic or nematic ordering occurring in a state bordering the FM
phase).  Nevertheless, such states with multipolar-ordering in the
zero-field case are deemed to be quite fragile.  Indeed, a recent
careful and accurate analysis of the spin-$\frac{1}{2}$ FM version of
the $J_{1}$--$J_{2}$ Heisenberg model (i.e., with $J_{1}<0$) on the
square lattice \cite{Richter2010:J1J2mod_FM}, which used both
high-order CCM and ED techniques, found that if such states did exist
(and no evidence at all was found for them in this study), they could
exist only over a very small range below $J_{2} \approx 0.4|J_{1}|$,
where the transition at which FM ordering disappears was accurately
determined to be $J^{c}_{2}=0.394(1)|J_{1}|$.  A recent Schwinger
boson study of the same model \cite{Feldner:2011_J1J2J3mod} also found
no evidence for any such states.

Similarly, in our present study of the checkerboard model we have
found no evidence at all to suggest that the CDVBC phase does not
extend all the way down to the FM phase.  Again, if any such
intermediate phase exists at all, it would need to be confined to a
very small range below $J_{2} \approx 0.7|J_{1}|$, on all the evidence
presented here, where the transition at which FM ordering disappears
is $J^{c_{3}}_{2}=0.66(1)|J_{1}|$.  On the other hand, the detection
of phases with novel quantum ordering, such as spin-nematic states, is
subtle, and the present checkerboard model may merit further
investigation in this respect in the very narrow region just above the
border of the FM phase. 

Finally, we note that there has been interest shown in frustrated
ferromagnets with respect to the formation of multimagnon bound states
under the influence of high magnetic fields (and see, e.g., \cite{Shannon:2006_FM,Kecke:2007_FM_magField,Sudan:2009_FM_magField,Nishimoto:2011_FM_magField}).
It might, therefore, be of interest to investigate the present
checkerboard model further, by coupling it to an external magnetic
field.  

\section*{Acknowledgment}
We thank the University of Minnesota Supercomputing Institute for the
grant of supercomputing facilities.

\section*{References}

\bibliographystyle{iopart-num}
\bibliography{bib_general}

\end{document}